\documentclass[aip,jcp,twocolumn,amsmath,amssymb,reprint,showpacs,footinbib,superscriptaddress,longbibliography,floatfix]{revtex4-1}

\usepackage[english]{babel}

\usepackage{dcolumn} 
\usepackage{tabularx}
\usepackage{bm}
\usepackage{graphicx}

\usepackage{braket}
\usepackage{amsfonts,amssymb,amsmath}
\usepackage{mathrsfs}
\usepackage{mathtools}
\usepackage{epstopdf}

\usepackage [autostyle, english = american]{csquotes}
\MakeOuterQuote{"}

\usepackage[colorlinks]{hyperref}
\hypersetup{%
        plainpages=true,
        breaklinks=true,
        hypertexnames=false,
        pageanchor=true,
        colorlinks=true,
        linkcolor={blue},
        citecolor={black},
        urlcolor={blue},
        pagecolor={black},
        anchorcolor={black}
      }

\usepackage{float}

\newcommand{\be}{\begin{equation}}
\newcommand{\ee}{\end{equation}}
\newcommand{\eqr}[1]{Eq.~(\ref{#1})}
\newcommand{\fig}[1]{Fig.~\ref{#1}}

\newcommand{\beq}{\begin{equation}}
\newcommand{\eeq}{\end{equation}}
\begin{document}
\title{Optimizing co-operative multi-environment dynamics in a dark-state-enhanced photosynthetic heat engine}

\author{Melina Wertnik}
\affiliation{Theoretical Quantum Physics Laboratory, RIKEN Cluster for Pioneering Research, Wako-shi, Saitama 351-0198, Japan}
\affiliation{Theoretical Physics, ETH Zurich, CH-8093 Zurich, Switzerland}

\author{Alex Chin}
\affiliation{Institut des NanoSciences de Paris, Sorbonne Universit\'e, 4 place Jussieu, boîte courrier 840, 75252 PARIS Cedex 05}

\author{Franco Nori}
\affiliation{Theoretical Quantum Physics Laboratory, RIKEN Cluster for Pioneering Research, Wako-shi, Saitama 351-0198, Japan}
\affiliation{Department of Physics, University of Michigan, Ann Arbor, Michigan 48109-1040, USA}

\author{Neill Lambert}
\email[e-mail:]{nwlambert@gmail.com}
\affiliation{Theoretical Quantum Physics Laboratory, RIKEN Cluster for Pioneering Research, Wako-shi, Saitama 351-0198, Japan}
\date{\today}
\pacs{}

\begin{abstract}
We analyze the role of coherent, non-perturbative system-bath interactions in a photosynthetic heat engine.
Using the reaction-coordinate formalism to describe the vibrational phonon-environment in the engine, we analyze the efficiency around an optimal parameter regime predicted in earlier works.
We show that, \textcolor{black}{in the limit of high-temperature photon irradiation}, the phonon-assisted population transfer between bright and dark states is suppressed due to dephasing from the photon environment, \textcolor{black}{ even in the Markov limit where we expect the influence of each bath to have an independent and additive affect on the dynamics.}
Manipulating the phonon bath properties via its spectral density enables us to identify both optimal low- and high-frequency regimes where the suppression can be removed.
This suppression of transfer and its removal suggests that it is important to consider carefully the non-perturbative and cooperative effects of system-bath environments in designing artificial photosynthetic systems, and also that manipulating inter-environmental interactions could provide a new multidimensional "lever" by which to optimize photocells and other types of quantum device.
\end{abstract}
\maketitle

\section{Introduction}

The observation of quantum coherent beating over time-scales comparable to energy transfer in a range of photosynthetic light-harvesting complexes \cite{chengreview,ishizakireview,Lambert_2013,Scholes2017} has spurred many approaches to identifying mechanisms by which coherence can increase efficiency of energy harvesting. For example, it has been proposed that the transport of excitation energy between a light-harvesting antenna and the reaction center in the Fenna-Matthews-Olsen (FMO) complex \cite{Engel2007,Akihito09,Collini2010} follows a wavelike coherent motion instead of hopping randomly from state to state \cite{Engel2007}, even at ambient temperatures \cite{Collini2010}.
In addition, noise \cite{Plenio08,RebentrostEtAlNJP2009,Caruso09,Plenio10,bleaching,Chin2013,Li2012,PhysRevE.92.052720} from the protein environment surrounding the complex  may also assist the energy transfer on its way, as it leads to line broadening and thermal excitation, which can increase the overlap between energetically separated states\cite{Chin2012,Huelga2013}.
Together these effects have been termed "environment assisted quantum tunneling", describing the interplay of quantum tunneling and thermal fluctuations to overcome energetic barriers to transport, not just in the FMO, but also in other complexes \cite{Engel2007,Collini2010,Lambert_2013,Scholes2017,romero2014quantum,fuller2014vibronic}.

Recently, another mechanism \cite{Dorfman_2013, Creatore_2013,Zhang_2015} was proposed, recasting the standard model of a light-absorbing photosynthetic reaction center as a "quantum heat engine".
The advantage of this new approach arises from two effects.
First, the collective dipole coupling of the two chromophores to the incident light leads to interference effects that create "bright" and "dark" states.
Other mechanisms, such as direct coupling\cite{Creatore_2013} or environment-induced interference effects\cite{Dorfman_2013}, can then lead to the energetic splitting of these states and subsequent breaking of detailed balance for the photon absorption process.
The suppression of spontaneous and stimulated emission of photons leads to an overall increased efficiency and can be thought of as a way to overcome the Shockley limit on absorption\cite{Shockley_1961}.
Secondly, the nature of the coherent and non-perturbative interaction between the dimer and its phonon environment has further potential to enhance the efficiency \cite{ Santamore2013,elinor1,Castro2014,killoran2015, Stones2016,Strasberg_2016, elinor2, Chen2016,Newman_2016, Qin2017}.
This is particularly appealing when such problems are recast as "quantum heat engines", as it allows us to view the problem of improving efficiency through the lens of non-Markovian thermodynamics and understand this phonon environment as a "cold bath", whose properties can be manipulated and taken advantage of to maximise energy extraction from a `hot' environment \cite{Strasberg_2016}.

The ultimate goal of these approaches is both to understand naturally existing systems and to identify clear mechanisms which enhance efficiencies of such "quantum light harvesting heat engines", which can then be applied to artificial photosynthetic systems and photocells \cite{bredas2017photovoltaic}. Indeed, understanding the physics of molecular open quantum systems has recently been highlighted as a key consideration for the design of a whole new class of `coherent' molecular devices with sensing and energy aplications \cite{Scholes2017}. In this work, we investigate how to modify the phonon environment to optimize the efficiency of such a light-harvesting "quantum heat engine", as realized by two coupled chromophores. To describe the complex problem of chromophores strongly interacting with their phonon (or vibronic) environment, which plays the role of the cold-bath in the heat engine picture, many techniques of varying degrees of sophistication and approximation have been developed.
These, among many \cite{deVega2017},  include perturbative but non-Markovian methods \cite{Chen2015, Fruchtman2016} and exact methods, like the "hierarchy equations of motion" \cite{Tanimura_1989,Akihito09,Kreisbeck2012}, the reaction-coordinate method\cite{Garg_1985,Iles_Smith_2014, Strasberg_2016, Iles_Smith_2015, Newman_2016}, and the quasi-adiabatic path-integral \cite{Makri_1998,Peter11} approach. For zero temperature simulations, powerful, numerically exact many-body methods are also availabe, such as the Multiconfigurational Time-Dependent Hartree Fock and Tensor Network  approaches which give access to the dynamics of both system and environment \cite{manthe2008multilayer, schulze2015explicit,chin2013role,prior2013quantum,schroder2017multi}.
The reaction-coordinate (RC) method, which we employ herein, has been proposed as a particularly powerful tool for these type of problems.
It allows for a straightforward evaluation of thermodynamic quantities \cite{Strasberg_2016, Newman_2016}, can be applied selectively to a many-bath system (in concert with normal Markovian methods to describe other baths), and can describe both low and high-frequency baths for arbitrary coupling strengths \cite{Ishizaki_2017}.

\subsection{Summary of results}

Here we show, using this reaction coordinate method, that in \textcolor{black}{the high-temperature photon irradiation limit, and fast donor-acceptor transfer, which in the classical case both optimizes the Carnot efficiency and leads to a large current,} the interaction between the chromophores and the phonon environment is dephased to such a degree that the photo-induced current is suppressed. \textcolor{black}{This result is an explicit example of how the interplay of multiple environments can be non-additive and lead to unexpected results \cite{Stones2016a,Giusteri2017, schroder2017multi}.} \textcolor{black}{Indeed, the importance of such effects extends beyond the case of light-harvesting to a wider range of nanoscale devices strongly coupled to multiple environments, such as in lead-dot(molecule)-lead structures\cite{gruss2017communication}.}
We show how this suppression can be overcome by increasing the coupling to the phonon environment such that it is more strongly influencing the chromophore system than an equivalent Markovian environment.

\textcolor{black}{The assumption of high-photon irradiation temperature \cite{Dorfman_2013, killoran2015, Creatore_2013, Zhang_2015, Chen2016} does not occur with natural sunlight.  Thus, we also examine, in \ref{sec:lowT}, the low-temperature photon irradiation case, with parameters corresponding to more realistic, but optimally concentrated, natural sunlight irradiation. Here we find that in the Born-Markov limit of the phonon environment, the baths again become additive.  We also show that entering into the non-Markovian and strong-coupling limit for this bath ultimately has diminishing returns, in this low-temperature case.  The mechanism of this diminishment is shown, via perturbation theory, to arise from the ``counter-rotating'' terms in the interaction with the phonons, which reduce the relative populations of the donor states which couple to the acceptor.}

The overall structure of this work is as follows. We begin by defining our heat engine model in section II.  We then describe the reaction-coordinate method for modelling the phonon environment in section III. In section IV we discuss the results of our model, and explore both high and low photon temperature regimes, and give our conclusions in section V. In the appendix we define the secular and non-secular Markovian master equations to which we compare our results.

\section{Model: Photosynthetic quantum heat engine}
\label{sec:model}
\begin{figure}
\centering
  \includegraphics[width=\columnwidth]{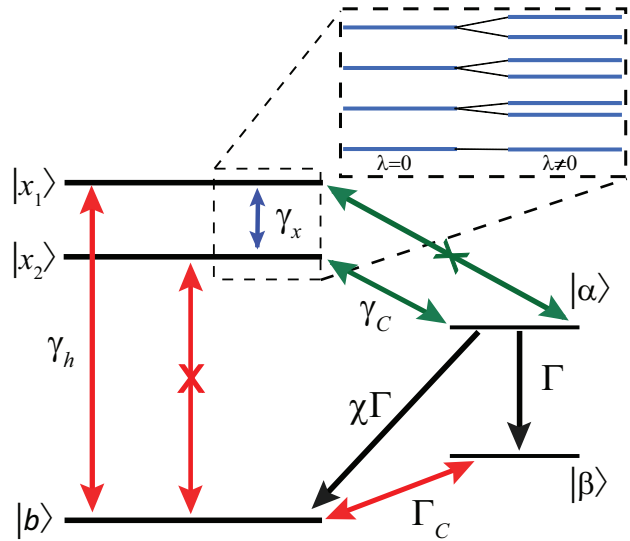}
 \caption{Scheme for the five level model \cite{Creatore_2013} in the coupled basis. A photon excites the donor molecules from the collective ground state $\ket{b}$, to the collective bright excited state $\ket{x_1}$. This excitation is then tranferred, via energy exchange with the phonon bath, to the dark state $\ket{x_2}$. Work can then be extracted via a charge transfer process to $\ket{\alpha}$ and $\ket{\beta}$ before the system is reset to the collective ground state. The red transition ($\gamma_h$) stems from absorption and emission of photons, the transitions due to transfer of electrons between donors and acceptor ($\gamma_c, \Gamma_c$) are rendered in green, while the transitions in black ($\Gamma, \chi\Gamma$) model the extraction of work as well as the recombination loss from $\ket{\alpha}$ to the ground state.
 With the Reaction Coordinate (RC) method the transition between the two donor excited states (given by a rate $\gamma_x$ in the Born-Markov model \cite{Creatore_2013}), is treated non-perturbatively by coupling the system to a collective degree of freedom in the phonon bath with coupling strength $\lambda_0$. The inset shows the lowest lying energy levels of the $x_1$ and $x_2$ states coupled to the collective RC mode for a coupling strength of $\lambda_0 = 1.06\times10^{-3}$ eV$^{3/2}$ at the resonant frequency $\Omega_0 = 0.03$ eV (cf. \fig{fig:2}a).}
 \label{fig:model}
\end{figure}

 A minimal model (cf. \fig{fig:model}) of a photosynthetic quantum heat engine describes the excitonic states of three molecules: a pair of coupled donor molecules $D_1$ and $D_2$ and one acceptor molecule $A$.
 As discussed by Creatore {\em et al.}  \cite{Creatore_2013}, a dipole-dipole interaction leads to strong coupling between the two donor molecules, transforming the originally degenerate excited states $\ket{a_1}$ and $\ket{a_2}$ into new non-degenerate eigenstates $\ket{x_{1,2}} = 1/\sqrt{2}(\ket{a_1}\pm\ket{a_2})$ with eigenvalues $E_{x_{1,2}} = E_{1,2} \pm J$, where $J$ is the coupling amplitude.
Due to the collective way the parallel dipole moments add up, only $\ket{x_1}$ is optically active.

Photon absorption (from a hot concentrated photon bath with thermal occupation $n_h$) occurs at a rate $\gamma_h$, causing transitions from the unexcited state $\ket{b}$ to the donor excited states $\ket{a_1}$ and $\ket{a_2}$.
Ultimately, the collective nature of that absorption leads to occupation of only the bright state $\ket{x_1}$, which then can relax to the dark state $\ket{x_2}$ by emission of phonons into an environment with thermal occupation $n_x$.
In the Markovian model this exchange is described by a rate $\gamma_x$, whereas in the RC model, which we describe in the next section, the interaction with a collective mode is used.
Occupation of the dark state breaks detailed balance, allowing time for an electron to be transferred to the acceptor molecule, taking the system to the acceptor excited state $\ket{\alpha}$, (at a bare rate $\gamma_c$, assisted by a thermal environment with occupation $n_c$).

This charge transfer is suppressed for the bright state due to the electron transfer matrix elements carrying opposing signs, $t_{D_1, A} = -t_{D_2, A}$.
At $\ket{\alpha}$, work is extracted, leading the system to the ground state of the acceptor molecule $\ket{\beta}$, at rate $\Gamma$.
Additionally, an effective lossy transition to the donor ground state is described by the rate $\chi\Gamma$, which represents additional losses by which the system is reset without producing work.
Finally, after work has been extracted, $\Gamma_c$ describes the other `reset' process of the system back to the ground state $\ket{b}$, e.g., due to donation of a charge to the original donor molecule from the environment \cite{Ghosh2009}.
In our model this transition is assisted by a different thermal environment with occupation $N_c$.
Note that all the transition rates involving the donor excited states are twice as large in the coupled basis than they are in the uncoupled basis.

The role of the phonon environment in transferring population must be considered carefully.
In real photosynthetic systems such environments typically exist in the non-perturbative and non-Markovian regimes, with both bath correlation times and system-bath correlation times being of the same order as excitonic couplings (the parameter $J$ in our model). Indeed, understanding the open system physics of this `intermediate regime' is the principle motivation for the diversity of powerful open system techniques mentions in the introduction.

The Hamiltonian describing the two donors, the acceptor molecule, and the phonon environment is
\begin{equation}
\begin{split}
H &= H_S + H_B + H_I\\
H_S &= \sum_{i=1}^5 E_i\ket{\psi_i}\bra{\psi_i} + J(\ket{a_2}\bra{a_1} + \ket{a_1}\bra{a_2})\\
H_B + H_I &= \frac{1}{2}\sum_k\Big[p_k^2 + \omega_k^2\Big(x_k - \frac{f_k}{\omega_k^2}s\Big)^2\Big],
\end{split}
\label{eq:ham}
\end{equation}
with mass-weighted positions $x_k = 1/\sqrt{2\omega_k}(c_k^{\dagger} + c_k)$ and momenta $p_k = i\sqrt{\omega_k/2}(c_k^{\dagger} - c_k)$ of the bath fulfilling $[x_k, p_l] = i\delta_{kl}$ (setting $\hbar\equiv1$ throughout) and with  creation and annihilation operators $c_k^\dagger$ and $c_k$.
The operator coupling system and bath is given by
\begin{equation}
s = \left(\ket{a_1}\bra{a_1}-\ket{a_2}\bra{a_2}\right),
 \end{equation}
 while the system is defined with the five states of the heat engine
 \begin{equation}
 \ket{\psi_i} = \left\{\ket{b}, \ket{x_1}, \ket{x_2}, \ket{\alpha}, \ket{\beta}\right\},
\end{equation}
 and corresponding energies $E_i$. Also, not shown here are the explicit Hamiltonians for the other baths shown in \fig{fig:model}.
Their influence on the model will be defined explicitly within the master equation picture later.

\section{Methods: The reaction coordinate}
\label{sec:methods}
The  dark-state-enhanced heat engine, including the dipole coupling $J$, has previously been studied \cite{Creatore_2013} using a Born-Markov master equation description of the phonon bath (see \eqr{SME} in the appendix), and been compared to a model where the donors do not couple to each other. That work  \cite{Creatore_2013} found a relative current enhancement of the coupled ($J>0$) over the uncoupled ($J=0$) case of roughly 30\% . To give a like-for-like comparison here, we primarily use the parameters found in \cite{Creatore_2013} that gave this large current enhancement  in our simulations (cf. Table \ref{tab:parameters}), \textcolor{black}{which focus on the limit of high photon irradiation.  In section \ref{sec:lowT} we will consider the low-temperature photon case, which is more applicable to systems irradiated by natural sunlight.}

\begin{table}
\caption{Parameters used for the numerical simulation}
\label{tab:parameters}
\begin{tabular}{p{0.4\linewidth}p{0.3\linewidth}p{0.3\linewidth}}
\hline\hline
Donor excitation energy&$\Delta_h = E_1-E_b$\newline $= E_2-E_b$ & 1.8 eV\\
Donor-acceptor \newline energy difference & $\Delta_c = E_1-E_\alpha$\newline$ = E_2-E_\alpha$
\newline $ = E_\beta - E_b$ & 0.2 eV \\
Excitonic coupling &$J$ & 0.015 eV\\
Photon absorption rate &$\gamma_{h}$ & $6.2 \times 10^{-7}$ eV \\
Donor-acceptor charge transfer rate &$\gamma_{c}$ & $6\times10^{-3}$ eV \newline (in \ref{sec:highT}), \newline $1\times 10^{-6}$ eV \newline (in \ref{sec:lowT})\\
Work extraction rate &$\Gamma$ & 0.124 eV  \\
Reset rate &$\Gamma_c$ & 0.0248 eV\\
Photon occupation & $n_h$ & $\sim 60000$ (in \ref{sec:highT}),  \newline  $\sim 0.03$ (in \ref{sec:lowT}) \\
Charge transfer \newline  bath occupation & $N_c = n_c $ at 300K & $\sim4.4\times 10^{-4}$\\
Phonon bath occupation \newline (for the Born-Markov model) & $n_x$ at 300K & $\sim0.46$ \newline  \\
\hline\hline
\end{tabular}
\newline\\
\end{table}

Instead of employing a Born-Markov master equation, we construct a more general master equation here, that can take into account non-Markovian effects of the phonon environment.
We do so using the reaction coordinate (RC) mapping \cite{Garg_1985} to transform the bath Hamiltonian such that a collective degree of freedom of said environment is included in the "system" Hamiltonian, while the residual environment is treated with a traditional generalized master equation.
The validity of this approach was carefully analyzed in \cite{Iles_Smith_2014, Iles_Smith_2015} and, by comparison to the hierarchy method, found to be exact.  Here we do not benchmark the results against another technique, as in those earlier works, and thus do not claim that our treatment of the phonon environment is exact to all orders for all parameter regimes we study herein. Rather, we propose that it captures more salient features than obtainable with a standard Born-Markov method alone.

\subsection{The reaction coordinate transformation}

We follow the description of the RC method in \onlinecite{Strasberg_2016} and repeat only the most salient steps here.
Starting from the complete Hamiltonian [\eqr{eq:ham}], a transformation has to be found that maps the interaction of the phonon bath with the system onto one new coordinate (the RC).
This reaction coordinate is itself coupled to a residual bath, which will be treated using the standard Born-Markov approximation.
However, later we will see that relaxation rates induced by the residual phonon bath, as well as the hot photon bath and the donor-acceptor transfer process, can be of the order of the system frequencies, such that we cannot make the secular approximation in describing these baths \cite{brumer2016}. Therefore, we work with Born-Markov non-secular master equations for these particular environments.

Under this transformation, the system can be coupled arbitrarily strongly to the RC, as this is treated exactly through the $H'_{RC}$ part of the transformed Hamiltonian
\begin{equation}
\begin{split}
H' &= H'_S + H'_{B} + H'_I\\
H'_S &= H_S + H_{RC}\\
H_{RC} &= \frac{1}{2}\Big[P_1^2 + \frac{g_0^2}{\delta\Omega_0^2}\Big(X_1 - \frac{\delta\Omega_0^2}{g_0}s\Big)^2\Big]\\
H'_B + H'_I &= \frac{1}{2}\sum_k\Big[P_k^2 + \Omega_k^2\Big(X_k - \frac{C_k}{\Omega_k^2}X_1\Big)^2\Big].
\end{split}
\label{eq:H_RC}
\end{equation}
Here $g_0$ is the coupling strength between the reaction coordinate and the system, while $X_1 = (a + a^{\dagger})/\sqrt{2\Omega_0}$ describes the collective quadrature of the RC mode.
In this new frame we treat the residual baths and their interaction with the system via the RC mode ($H'_B + H'_I$) with the non-secular Born Markov master equation.
The renormalization term $\delta\Omega_0^2 s^2/2$ can be neglected as it is small compared to the energy difference between the excited states of the dimer and the states $\ket{b}$ and $\ket{\alpha}$.

The details of the transformation of the Hamiltonian above are entirely defined by the choice of spectral density for the {\em original} phonon environment,
\begin{equation}
J_0(\omega) \equiv \frac{\pi}{2}\sum_k\frac{f_k^2}{\omega_k}\delta(\omega-\omega_k).
\end{equation}
Here we focus on the underdamped Brownian oscillator spectral density,
\begin{equation}
J_0(\omega)= \frac{\lambda_0^2\gamma\omega}{(\omega^2 - \Omega_0^2)^2 + \gamma^2\omega^2},
\label{eq:SD_0}
\end{equation}
as it allows us to investigate the influence of resonant and off-resonant structured environments that can dominate the vibrational spectrum of light-harvesting complexes \cite{wendling}.

The choice \cite{Strasberg_2016} of spectral density ultimately fixes the coupling strength $g_0$ and the RC frequency $\frac{g_0}{\delta\Omega_0}$.
From the transformation of the Hamiltonian we obtain a term for the coupling strength as well as for the system renormalization:
\begin{equation}
\begin{split}
g_0^2 &= \sum_k f_k^2\\
\delta\Omega_0^2 &= \sum_k \frac{f_k^2}{\omega_k^2}.\\
\end{split}
\end{equation}
These can then be expressed through integrals over the spectral density $J_0(\omega)$ and evaluated as
\begin{equation}
\begin{split}
g_0^2 &=  \frac{2}{\pi}\int_0^\infty d\omega J_0(\omega)\omega = \lambda_0^2\\
\delta\Omega_0^2 &= \frac{2}{\pi} \int_0^\infty d\omega \frac{J_0(\omega)}{\omega} = \frac{\lambda_0^2}{\Omega_0^2}.\\
\end{split}
\end{equation}
The physical frequency of the RC is given by $\frac{g_0}{\delta\Omega_0} = \Omega_0$ and we can now identify the coupling strength $g_0 = \lambda_0$.

The spectral density of the residual bath is connected to the original environment through a recursion relation \cite{Martinazzo_2011}.
By deriving the Fourier space propagator from the original and transformed Hamiltonian and comparing them \cite{Strasberg_2016}, the transformed spectral density can be found by evaluating
\begin{equation}
J_1(\omega) = \frac{\lambda_0^2J_0(\omega)}{|W_0^+(\omega)|^2},
\end{equation}
where
\begin{equation}
W_0^+(z) = \underset{\epsilon\rightarrow 0^{+}}{\mathrm{lim}}\frac{1}{\pi}\int_{-\infty}^{\infty}d\omega\frac{J_0(\omega)}{\omega-(z+i\epsilon)}
\end{equation}
is the Cauchy transform of the original spectral density $J_0$.
Using the residue theorem, our choice of the original spectral density [\eqr{eq:SD_0}] leads to an Ohmic spectral density for the residual bath, $J_1(\omega) = \gamma\omega$.

\subsection{Master equation approximations}
We are mainly interested in the steady-state current output for various input parameters, which is calculated numerically from the steady-state density matrix $\rho(t\rightarrow \infty)$ \cite{qutip, qutip2}.
The master equation used to solve for $\rho(t\rightarrow \infty)$ is split into three parts:
\begin{enumerate}
\item The transition between the donor excited states, $\ket{x_1}$ and $\ket{x_2}$, due to the interaction with the phonon environment, is treated by the system and RC parts of the Hamiltonian $H'_S$.
\item The residual baths coupled to the RC mode, as well as the transitions involving the donor excited states, are described by a non-secular Born-Markov master equation. \label{item:residual_bath}
\item The transitions not involving the donor excited and RC states are treated using phenomenological secular Born-Markov Lindblad generators.
\end{enumerate}

We cannot use the secular approximation in point \ref{item:residual_bath}, because, to do so, the timescales of the system have to be much smaller than the relaxation times: $\tau_S \ll \tau_R$.
The system timescale is proportional to the inverse of the energy splitting, while the relaxation time is defined through the transition rates and occupation numbers.
The condition for the secular approximation to be valid then becomes:
\begin{equation}
2J, \frac{\lambda_0}{\sqrt{2\Omega_0}} \gg (n_{h}+1)\gamma_{h}, (1+n_c)\gamma_c, (1+n_x)\gamma.
\end{equation}
Even without the RC-mode description of the phonon bath, this condition is not fully satisfied because the rate of absorption and emission of photons, $(n_{h}+1)\gamma_{h}$, is comparable to $2J$.
However, by introducing the RC, the energy splittings in the system Hamiltonian become even smaller (cf. \fig{fig:model}) such that a non-secular approach is necessary for both the photon bath as well as the electron transfer.

Our general non-secular master equation is

\begin{equation}
\begin{split}
\dot{\rho}(t)&=-i[H'_S, \rho(t)] + \mathcal{M}_{X_1}[\rho(t)]+ \mathcal{M}_{Q_h}[\rho(t)]+ \mathcal{M}_{Q_c^{(\alpha)}}[\rho(t)] \\
&+ \mathcal{L}_{\beta b}[\rho(t)]+ \mathcal{L}_{b\beta}[\rho(t)] + \mathcal{L}_{\alpha\beta}[\rho(t)]+ \mathcal{L}_{\alpha b}[\rho(t)],
\end{split}
\label{RCME}
\end{equation}
where the non-secular generators \cite{Breuer2002} are
\begin{equation}
 \mathcal{M}_{A}[\rho(t)] = - [A, [\chi, \rho(t)]] + [A, \{\Theta, \rho(t)\}],
\end{equation}
with the system operator $A$ which couples to the given environment and
\begin{equation}
\begin{split}
\chi &= \frac{1}{2}\sum_{kl}J_A(\omega_{kl})\mathrm{coth}\Big(\frac{\beta\omega_{kl}}{2}\Big)A_{kl}\ket{k}\bra{l}\\
\Theta &= \frac{1}{2}\sum_{kl}J_A(\omega_{kl})A_{kl}\ket{k}\bra{l}.\\
\end{split}
\label{MQ}
\end{equation}
For the residual bath the system operator is given by the RC mode $X_1=(a + a^{\dagger})/\sqrt{2\Omega_0}$ and the spectral density by $J_1(\omega)$.
For the transitions between $\ket{b}$ and $\ket{x_1}$ and between $\ket{x_2}$ and $\ket{\alpha}$ the transition operators
\beq
Q_h = (\ket{a_1} + \ket{a_2})\bra{b} + \ket{b}(\bra{a_1}+\bra{a_2}),
\eeq
and
\beq
Q_c^{(\alpha)} = \ket{\alpha}(\bra{a_1}-\bra{a_2}) + (\ket{a_1} - \ket{a_2})\bra{\alpha},
\eeq
 are used and the spectral density for these baths is assumed to be Ohmic,
\begin{equation}
J_{h,c}(\omega) = \frac{\gamma_{h,c}}{2\Delta_{h,c}}\omega,
\end{equation}
for each of them respectively, \textcolor{black}{ where the donor excitation energies are $\Delta_{h} = E_1 - E_b = E_2 - E_b$, and the donor-acceptor energy difference is $\Delta_{c} = E_1 - E_{\alpha}$, as per Table \ref{tab:parameters}.}
The eigenvectors $\ket{k}$ and eigenenergies $\omega_{kl}$ are calculated from $H'$.
The residual phonon bath temperature and donor-acceptor transition rate temperature is set to $300$ K, but the transition between the ground state and $\ket{x_1}$ is governed by the photon bath, which operates at a much higher temperature, defined through its occupation number given in Table \ref{tab:parameters}.

The transitions which do not couple directly to the excited donor states are given by secular Born-Markov Lindblad generators
\beq
\mathcal{L}_{AB}= \left[C_{AB}\rho(t)C_{AB}^{\dagger}- \frac{1}{2} \left\{C_{AB}^{\dagger}C_{AB},\rho(t)\right\}\right],
\eeq
 with
\begin{equation}
\begin{split}
C_{\beta b} &= \sqrt{\Gamma_c N_c}\ket{\beta}\bra{b}\\
C_{b\beta} &= \sqrt{\Gamma_c(1+N_c})\ket{b}\bra{\beta}\\
C_{\alpha\beta} &= \sqrt{\Gamma}\ket{\beta}\bra{\alpha}\\
C_{\alpha b} &= \sqrt{\chi\Gamma}\ket{b}\bra{\alpha}.\\
\end{split}
\label{c_ops}
\end{equation}

The two equivalent (secular and non-secular) Markovian master equations are given in the appendix.


\section{Results and discussion}
\label{sec:highT}

\begin{figure*}
 \includegraphics[width=.48\linewidth]{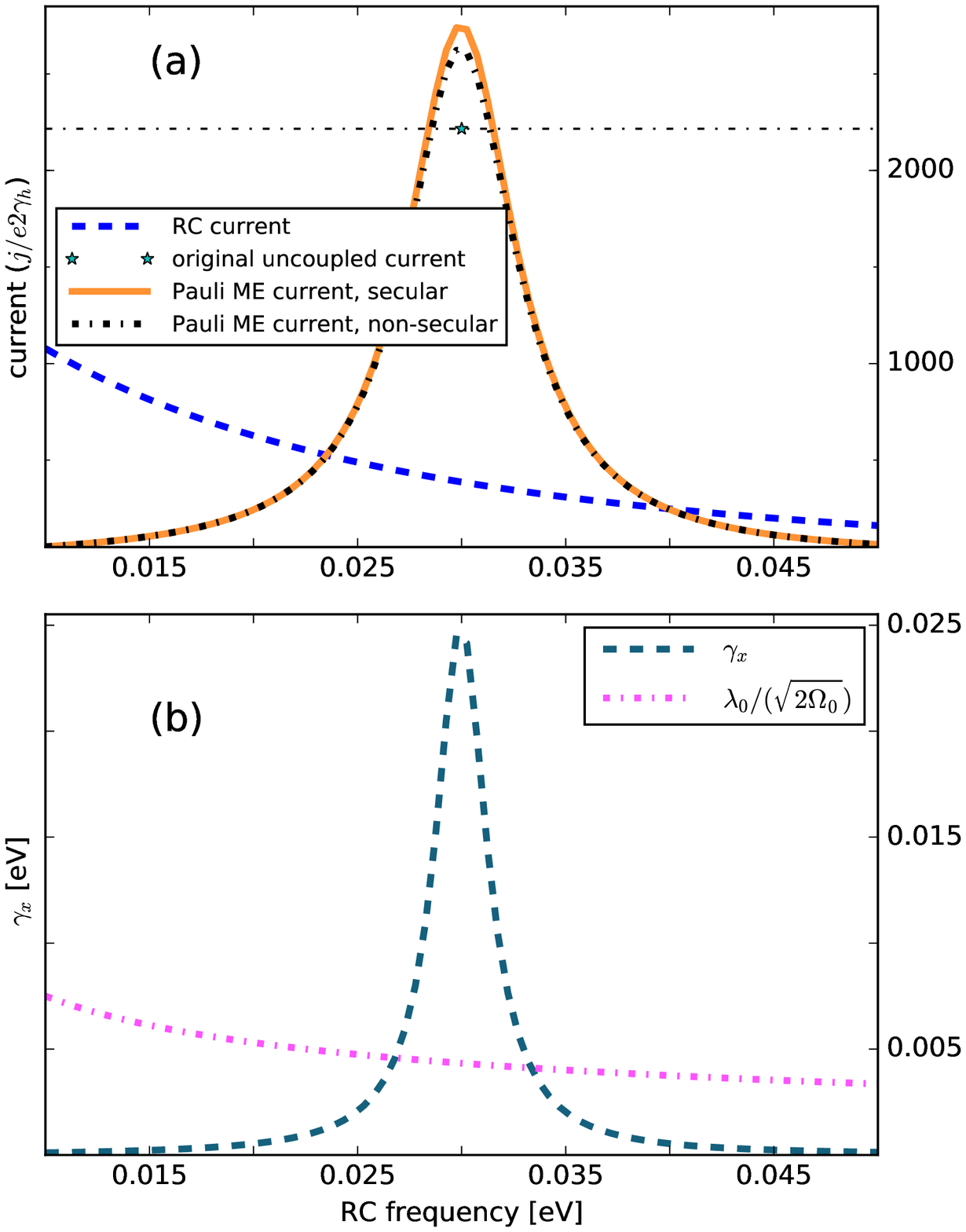}
  \includegraphics[width=.48\linewidth]{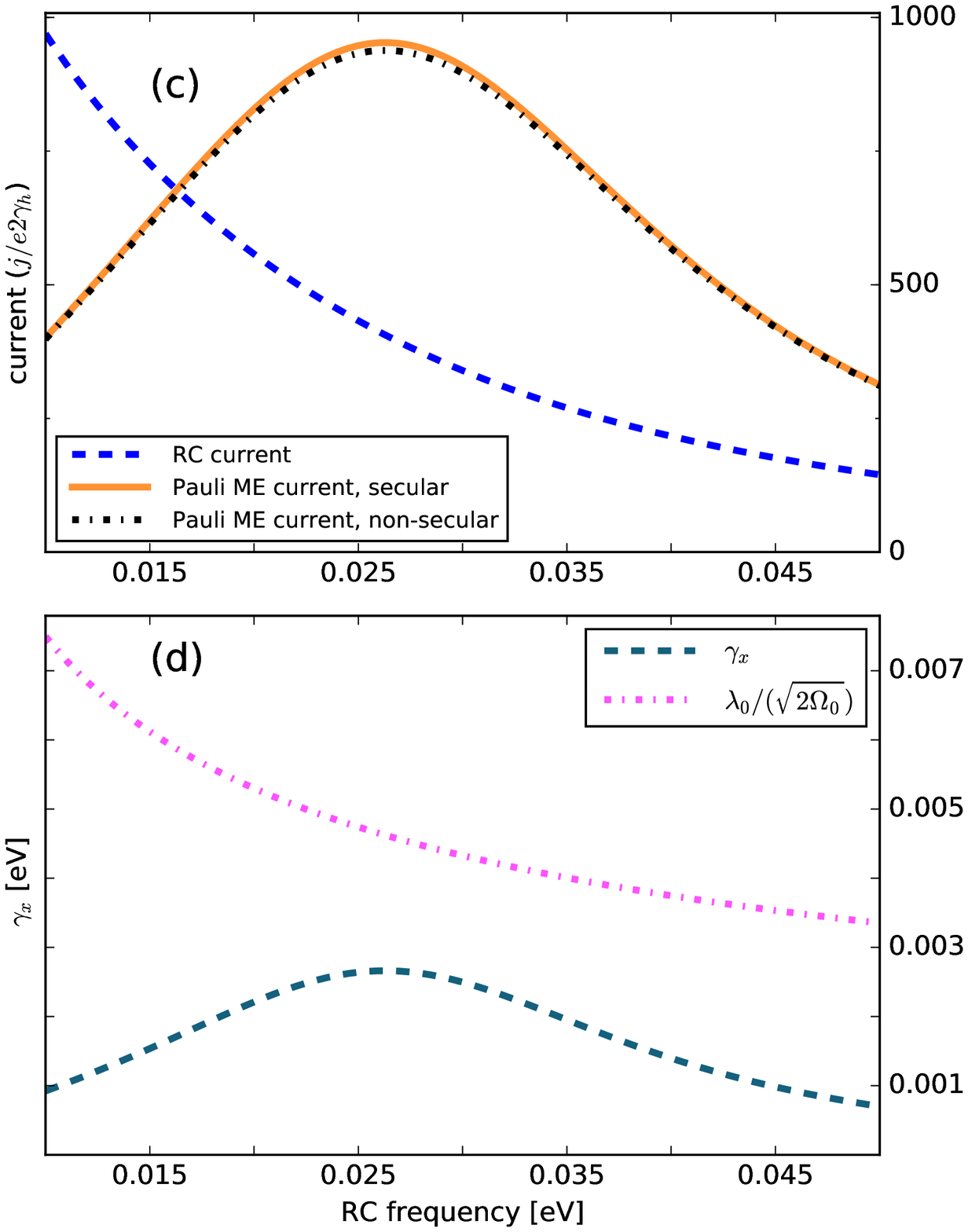}
 \caption{Current as a function of the reaction coordinate (RC) frequency $\Omega_0$ [eV] (i.e., the peak in the spectral density), for $\lambda_0=1.06\times10^{-3}$ eV$^{3/2}$, and for different values of $\gamma$.
This is compared to the current for a model without the RC-mode, which has been calculated using the equivalent Born-Markov master equation, and is a function of the transition rate $\gamma_x$ corresponding to the given RC frequency and coupling strength.
(a) $\gamma=0.1\Omega_0$: In this example we use the default parameters shown in Table \ref{tab:parameters} and compare the current from the secular (orange curve) and non-secular Born-Markov ME (red dash-dotted curve) and the RC method (blue dashed curve).  The magnitude of the RC current at resonance only reaches about $15$\% of the one from the Born-Markov ME. At lower frequencies the RC model predicts an increase in the current, which surpasses the equivalent Born-Markov ME current. As a reference point, the blue star/horizontal dashed line is the value of the current for the uncoupled model, $J = 0$, also calculated using an equivalent Born-Markov master equation, but with no phonon bath at all. Figure (b) shows, with a blue-dashed curve, the dependence of $\gamma_x$ on the resonant frequency (RC frequency) of the phonon spectral density. The pink dashed-dotted curve shows the effective coupling strength $\lambda_0/\sqrt{2\Omega_0}$ between system and RC mode.  (c) $\gamma=\Omega_0$: Using the same coupling strength $\lambda_0$ but changing the broadness parameter of the spectral density leads to a decreased transition rate $\gamma_x$, thus the current from the Born-Markov ME lowers and broadens in turn. In this case, the RC current remains mostly unchanged, as it follows the behavior of the effective coupling strength [shown in (d)] instead of the transition rate [also illustrated in (d)].}
\label{fig:2}
\end{figure*}

Here we will present and discuss the characteristics of the current generated by our RC method and compare them with the secular Born-Markov master equation model studied in \cite{Creatore_2013}, c.f., \eqr{SME}, as well as the non-secular equivalent, \eqr{NSME}, which takes into account that the transitions $\gamma_h n_h$ and $\gamma_c n_c$ can be of order of the energy splitting $2J$.
We define the current as proportional to the rate of population transfer from  $\ket{\alpha}$ to $\ket{\beta}$, i.e.,
\begin{equation}
j = e \Gamma \bra{\alpha}\rho \ket{\alpha}.
\end{equation}
We choose an effective transition rate for the influence of the phonon environment on the system such that $\gamma_x \equiv 2J_0(\omega=2J)$ which, after setting the resonance  $\Omega_0$ and width $\gamma$, determines $\lambda_0$ in the RC model.
For example, choosing $\gamma_x = 25$ meV, and setting the spectral density to be resonant $\Omega_0 = 2J$, and choosing a narrow distribution, $\gamma=0.1 \Omega_0$, $\lambda_0 = \sqrt{J\gamma\gamma_x}$,  we find, surprisingly, that the current is roughly three times lower using the RC method than using either Born-Markov master equations (cf. \fig{fig:2}).

Away from resonance, the relaxation rate $\gamma_x$ decreases rapidly.
As mentioned, the spectral density has been chosen such that the value for $\omega = \Omega_0$ corresponds to the transition rate $\gamma_x$.
Thus one would expect a maximal current at resonance, where the frequency of the phonon bath corresponds to the energy gap between states $\ket{x_1}$ and $\ket{x_2}$ and $\gamma_x$ is maximal for constant coupling strength $\lambda_0$.
If one increases the RC frequency above the resonance frequency, the current indeed decreases steadily towards zero, albeit this decay is much slower in the RC method than in the Born-Markov master equation.

\subsection{Suppression of Current}
The counter-intuitive suppression of the current arises from the non-secular nature of the {\em photon} environment and is not captured by a standard secular master equation.
Its origin lies in the fact that the strong illumination leads to fast emission and absorption between $\ket{x_1}$ and $\ket{b}$, such that phonon-mediated coherence between $\ket{x_1}$ and $\ket{x_2}$ cannot be built fast enough.
The interaction with the phonon bath is effectively dephased by the high temperature of the photon bath, and population transfer to the $\ket{x_2}$ state cannot occur.
In other words, the photon environment rapidly "measures" the population in $\ket{x_1}$, causing a quantum Zeno effect.
This is akin to the suppression of lasing seen in single-atom lasers at high inversion rates \cite{ashhab}.
 A minimal condition for this "Zeno" effect is,
\begin{equation}
\gamma_h n_h, \gamma_c \gg \frac{\lambda_0}{\sqrt{2\Omega_0}}.
\label{eq:criterion}
\end{equation}

  Using a non-Markovian bath under such strong illumination then seems, at first glance, counter-productive. Off-resonance  we can identify regions where the current exceeds that
predicted by the equivalent Markovian master equation (cf. \fig{fig:2}), but these remain smaller than those predicted by the uncoupled dimer model. In this parameter regime, the current from the RC model predominantly follows the change in the effective coupling strength $\lambda_0/\sqrt{2\Omega_0}$, increasing for lower RC frequencies and decreasing for higher ones, due to the corresponding dependence of the maxima in the original spectral density.  In addition, because we scale the width of the bath while changing the resonance frequency, $\gamma\propto \Omega_0$, at lower frequencies the phonon environment is correspondingly more non-Markovian.

\begin{figure}
\includegraphics[width=\linewidth]{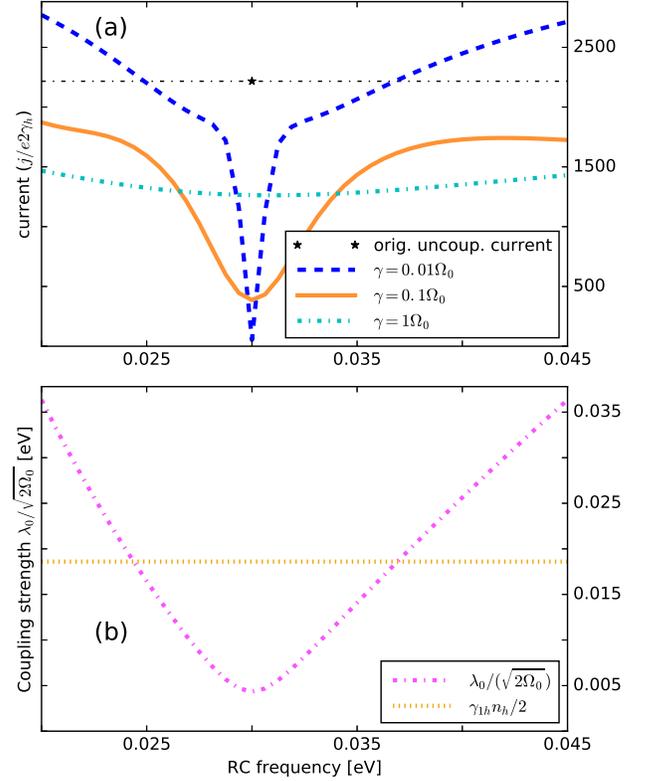}
\centering
\caption{(a) Current as a function of RC frequency $\Omega_0$ [eV], where the transition rate $\gamma_x$ is fixed, so that the coupling strength $\lambda_0$ increases as $\Omega_0$ is moved off resonance. Both the default spectral density parameter $\gamma=0.1\Omega_0$ and the sharper one $\gamma = 0.01\Omega_0$ lead to a sharp dip in current around resonance, the onset of which occurs around $\lambda_0/\sqrt{2\Omega_0}\approx \gamma_h n_h/2$, whereas the broader one $\gamma=\Omega_0$ is much flatter, and exhibits no dip. (b) The behavior of the coupling strength  $\lambda_0/\sqrt{2\Omega_0}$ (dot-dashed pink curve) for $\gamma=0.1 \Omega_0$,  illustrates that the onset of the dip in the current does occur around  $\lambda_0/\sqrt{2\Omega_0}\approx \gamma_h n_h/2$ (dashed orange curve). }
\label{fig:gamma_range}
\end{figure}

More generally, one expects that the current should be restored if the coupling to the phonon-environment is made competitive with the influence of dephasing, per \eqr{eq:criterion}.   \textcolor{black}{ In \fig{fig:low_photon_T_L}b we can see that increasing the coupling strength between system and RC when they are resonant can increase the current, overcoming this suppression, but this current is still less than that predicted by the Born-Markov master equation, though it can exceed the uncoupled dimer model for exceedingly strong coupling. To find a regime where the RC model exceeds the Born-Markov master equations and the uncoupled dimer model we} simultaneously vary both RC frequency $\Omega_0$ and coupling strength $\lambda_0$, to keep the effective rate that the system sees in the Markovian model $\gamma_x \equiv 2J_0(\omega=2J)=25$ meV fixed. In this case, we find a pronounced minimum around the resonant RC frequency (cf. figure \ref{fig:gamma_range}), again because of the Zeno-like effect seen in \fig{fig:2}.
Under these constraints, for both lower and higher frequencies, the effective coupling strength increases away from resonance and so does the current, as the energy splittings due to interaction with the phonons exceed the suppressive dephasing rates.

In particular, one can immediately note that a very narrow spectral density ($\gamma = 0.01 \Omega_0$) can lead to a substantial current off-resonance.
At very high frequencies (not explicitly shown in the figure) this can exceed the current predicted by the Born-Markov master equation for this optimal rate.   \textcolor{black}{  This suggests that there are regimes where a structured environment can enhance the current.
However, further optimizing this current, beyond the range shown in the figure, requires us to enter regime that exceeds the limits of our numerical RC method, and may be more amenable to methods used in other works \cite{Qin2017,Chen2016}. } In addition, it should be noted that, as we tune far off resonance in figure \ref{fig:gamma_range}, for the very narrow bath $\gamma = 0.01 \Omega_0$, numerical convergence in the phonon Fock-space becomes difficult, and the observed current should only be considered as a lower bound.

The dependence on the width of the spectral density in figure \ref{fig:gamma_range} also illustrates that while a sharper spectral density suppresses the current at resonance almost completely, and allows for a large current when off-resonance, a broader spectral density, with a correspondingly stronger coupling on resonance (to maintain the same $\gamma_x$ effective rate), results in a larger current on resonance (albeit one still below that predicted by the Markov theory).
When the same parameter $\gamma$ is changed, but the coupling strength is kept constant (cf. \fig{fig:2}c), the RC current does not change significantly, while the equivalent Born-Markov master equation current falls considerably (as it follows the correspondingly slower transition rate $\gamma_x$). This suggests it is predominantly the coupling strength itself, not the width of the spectral density, that increases the current in the Zeno-regime.  As the coupling strength is increased so that we leave the Zeno regime, the width of the spectral density begins to have a larger influence, as per  figure \ref{fig:gamma_range}.

\begin{figure}
\includegraphics[width=1.05\linewidth]{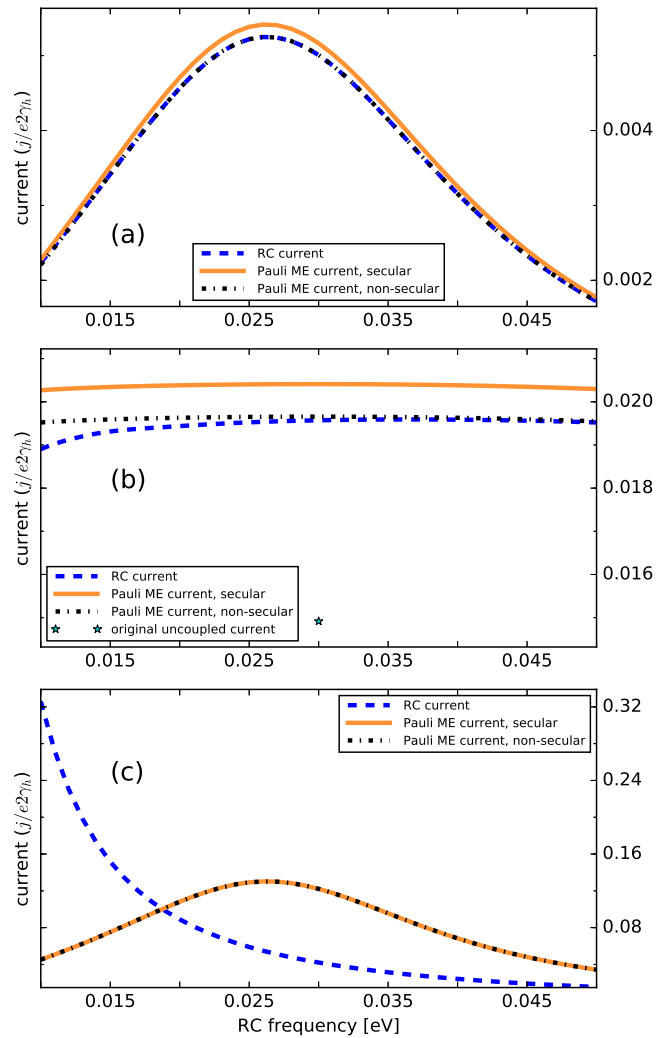}
\centering
\caption{(a) Current as a function of RC frequency [eV] for a fixed weak coupling strength $\lambda_0=1.06\times10^{-5}$ eV$^{3/2}$, with a correspondingly reduced photon temperature such that $n_h=0.03$, and reduced donor-acceptor rate $\gamma_c = 1\times 10^{-6}$, and for a broad phonon bath $\gamma=\Omega_0$. In this limit we see that the non-secular master equation and the RC model predict the same current, suggesting the phonon environment is in a weak-coupling Markov limit, and that all environments are additive. (b) For the same low photon intensity $n_h=0.03$, and donor-acceptor rate $\gamma_c = 1\times 10^{-6}$,  we see that increasing the phonon-emission rate back to $\gamma_x = 25$ meV, by raising the phonon coupling to $\lambda_0=1.06\times10^{-3}$ eV$^{3/2}$, and choosing $\gamma = 0.1 \Omega_0$, leads to a larger current, which can surpass that given by uncoupled dimers for equivalent parameters.  We also see that the Markov models tend to overestimate the current.
(c) When the phonon environment is in the weak-coupling-Markov limit ($\lambda_0=1.06\times10^{-6}$ eV$^{3/2}$, $\gamma= \Omega_0$), but the photon temperature is large $n_h=60,000 $ and the donor-acceptor rate is large $\gamma_c = 6\times 10^{-3}$, we see again that the baths become non-additive, even when each individual bath could be considered to be in the Born-Markov limit. }
\label{fig:low_photon_T}
\end{figure}

\subsection{Weak photon illumination regime and the Markov limit}
\label{sec:lowT}
\textcolor{black}{The assumption of a very-high photon temperature, employed in many earlier works on this heat-engine model \cite{Dorfman_2013, killoran2015, Creatore_2013, Zhang_2015, Chen2016}, maximizes the Carnot efficiency, and, in the classical model, maximizes the current.  However it does not  occur in natural or artificial photosynthetic systems, or photocells.  In addition, as we showed in the previous section, it can lead to suppressed current flow due to dephasing. In \onlinecite{amir2015} the authors consider the case of low-temperature illumination of a similar photosynthetic reaction center heat engine, corresponding to an maximal concentrated natural sunlight temperature of $6000$ K, and hence a thermal occupation of $n_h =0.03$, while also including the influence of non-degenerate dimer energies.  In such conditions they still found an advantage to the mechanism of including dimer-dimer couplings and subsequent bright to dark state conversion.  Here, we can also consider the influence of the RC mode at lower illumination levels, and slower donor-acceptor transfer rate $\gamma_c$ (larger values of $\gamma_c$ were found to be a reduce the current enhancement, relative to the uncoupled dimer example, in \onlinecite{amir2015}). This also allows us to finally answer the question of what happens in the weak phonon-coupling and Markov limit.}

\textcolor{black}{In \fig{fig:low_photon_T}a we observe that simultaneously lowering the temperature of the photon bath, donor-acceptor transfer rate $\gamma_c$, and the phonon coupling strength $\lambda$, and choosing a broad phonon spectral density, restores the Born-Markov limit, such that all baths become additive, and all three models begin to coincide (RC, secular and non-secular master equations). Raising the phonon-coupling, and choosing $\gamma$ so that the effective rate $\gamma_x=25$ meV, as in the previous section, we see in  \fig{fig:low_photon_T}b that the current again increases, and exceeds the uncoupled dimer model current. The dependence on resonance now is very weak, because of the bottleneck from the other rates. In this case the Born-Markov master equations tend to overestimate the current, and the full RC model predicts a slightly smaller current than expected.}

\begin{figure}
\includegraphics[width=1.05\linewidth]{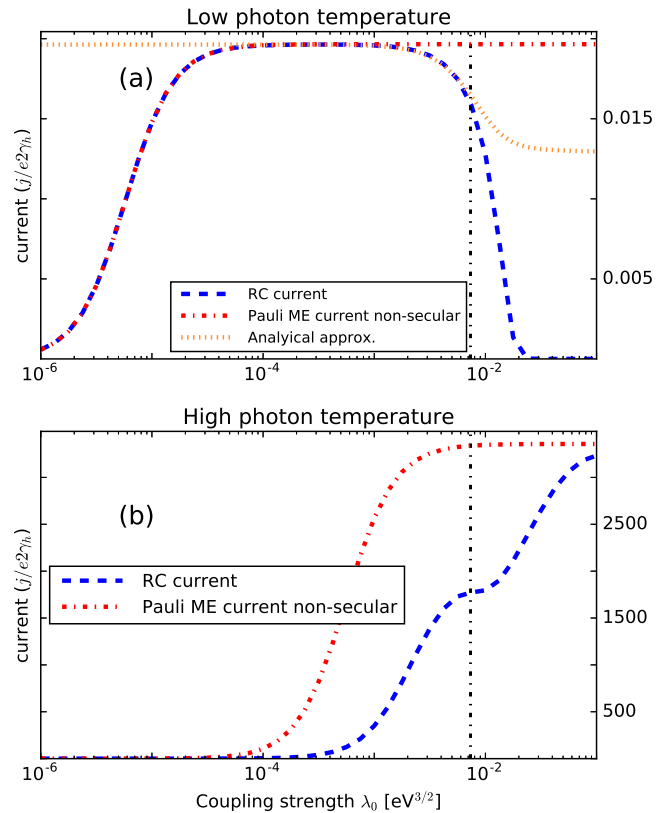}
\centering
\caption{ Current as a function of coupling strength $\lambda_0=$ with the RC on resonance $\Omega_0 = 2J$. (a) is for the low-photon temperature such that $n_h=0.03$, and reduced donor-acceptor rate $\gamma_c = 1\times 10^{-6}$, and for $\gamma=0.1\Omega_0$. Both the Pauli master equation and the RC model predict an optimal current around $\lambda = 10^{-3}$ eV$^{3/2}$, because the current is limited by the weak photon illumination rate and donor-acceptor rate. For strong couplings, the Pauli master equation over-estimates the current, because in the full RC model the strong interaction with the phonons reduces the matrix-elements which lead to current flow (see main text). The orange dotted curve shows an approximate expression explaining the quadratic dependence of the current on the coupling strength. (b) shows the current dependance on coupling for the high-temperature limit $n_h=60000$, and fast donor-acceptor rate $\gamma_c = 6\times 10^{-3}$. As in earlier figures we see that the RC current is suppressed with respect the Pauli master equation, due to the effects described in the main text.  The RC model predicts an inflection as the coupling strength becomes comparable with the resonance frequency (the line $\lambda_0=\Omega_0^{3/2}\sqrt{2}$ is plotted a dashed vertical line).  }
\label{fig:low_photon_T_L}
\end{figure}

\textcolor{black}{If we directly increase the coupling strength between dimer and RC on resonance, \fig{fig:low_photon_T_L}a, we eventually see diminishing returns in the RC model, with the current decreasing at a rate proportional to $\lambda_0^2$ as we enter a regime where $\lambda_0 > 10^{-3}$ eV$^{3/2}$. This is at first glance counter-intuitive;  however, if we consider that the time-scales of the excited states of the dimer interacting with the RC are much faster than the other rates in the model, we notice that the thermal state of the dimer and RC subspace begins to dominate the behavior (the excited states and the RC thermalize before a donor-acceptor transition can occur).  In that case, the transition to the acceptor state $\alpha$ will follow the transition matrix elements of the eigenstates of the dimer and RC subspace, weighted by their thermal populations. At $300$ K this is dominated by the first three excited states of the subspace. We can express these states perturbatively in the renormalized coupling strength $\chi =  \lambda_0/(\sqrt{2}\Omega_0^{3/2})$, which, for example, for the ground state, gives, when $\Omega_0=2J$,}
\begin{equation}
\ket{G} = \left(1-\frac{\chi^2}{8}\right)\ket{x_2,0}-\frac{\chi}{2}\ket{x_1,1} + \frac{\chi^2}{2\sqrt{2}}\ket{x_2,2}.
\end{equation}
\textcolor{black}{  The matrix elements connecting these states to the acceptor arise through the $\ket{x_2,n}$ contributions. Interestingly, some of these contributions, like the reduction in probability of being in the  $\ket{x_2,0}$ part of the ground state, arise due to the counter-rotating terms in the interaction with the phonons, and leads to a corresponding reduction in the current proportional to $1-\frac{\chi^2}{4}$.  Summing up such matrix elements for all eigenstates of the dimer-RC subspace, and weighting them by their thermal occupations, provides a way to check the validity of this assumption, with an approximate expression for the current}
\begin{eqnarray}
\bar{j}(\chi)/2e\gamma_h  &\approx& \mathrm{max}_{\chi}[j(\chi)/2e\gamma_h] \times \\
&&\sum_{k} P_{k}\left(\sum_n |\langle x_2,n|\psi_k\rangle |^2\right)/\mathcal{N} \nonumber
\end{eqnarray}
\textcolor{black}{The eigenstates $\psi_k$ are calculated numerically for the reduced model of the dimer excited states and RC from the Hamiltonian $H''_{S}$, which is $H'_{S}$ projected onto the subspace containing just the dimer states $x_1$, $x_2$, and the RC states. The thermal occupation probabilities $P_k$, are calculated from the thermal distribution $\rho =\exp(- H''_{S}/k_{\mathrm{B}}T)/Z$, where $Z=\mathrm{Tr}[\exp(- H''_{S}/k_{\mathrm{B}}T)]$. The normalization of $\bar{j}(\chi)$ by $\mathcal{N}$ is chosen so that in the limit $\chi\rightarrow 0$, $\sum_{k} P_{k}\left(\sum_n |\langle x_2,n|\psi_k\rangle |^2\right)/\mathcal{N} = 1$.  This approximate current is plotted in \fig{fig:low_photon_T_L}a, and agrees well with the full numerical current of the RC model for $\lambda_0$ in the range of applicability (a explicit second-order perturbative expansion also fits well for a smaller range of $\chi$, but we do not explicitly show it here because it is unwieldy). The approximate expression and the full numerics exhibit deviations for large $\chi$ (and of course for for very small $\chi$, where the current actually falls to zero, as there are no transitions to $\ket{x_2}$ states at all in this limit). }

\textcolor{black}{Finally, in \fig{fig:low_photon_T}c we again set the phonon-bath parameters to be the same as \fig{fig:low_photon_T}a, but raise $\gamma_c$ and the photon temperature back to their values used in the previous section. As in the previous section, we again see a large discrepancy in the current, with a suppressed value for the RC model;  this suggests that the issue of non-additive baths is more general than highly ``non-Markovian'' regimes, as is often assumed.}


\section{Conclusion}

In this work, we analyzed the influence of a non-Markovian phonon bath on the efficiency of light harvesting in a photosynthetic reaction center considered as a quantum heat engine.
Inspired by earlier works, which found an advantage in breaking detailed balance by inducing energetic splitting between bright and dark states, we modeled the phonon-mediated population transfer process from bright to dark states with the reaction coordinate formalism.
This formalism allows us to model the influence of phonons for a wide range of coupling strengths and bath memory times.

We found that, counter-intuitively,  bright photon illumination can suppress the current in the heat engine, due to quantum Zeno-like dephasing of the exciton-phonon interaction.
This suppression can be avoided by tuning properties of the phonon bath, and in doing so one can even generate currents significantly larger than the equivalent Markovian model.  In the low photon temperature limit, in contrast, increasing the coupling to the phonon environment tends to reduce the current.
Our results suggest that non-Markovian environments can be used to enhance the efficiency of light-harvesting systems, but that the properties of the environment must be chosen, or designed, in artificial systems, carefully, as their influence can also be detrimental, in certain situations.

\textcolor{black}{In addition, our results suggest that care must be taken whenever one considers multiple environments coupled to a single system, particularly when those environments induce dynamics on very different time scales \cite{Giusteri2017}, even when one is apparently in a weak-coupling Markov limit for all baths}. This is a potentially useful observation, as it suggests that dissipative processes due to many commonly encountered perturbative environments, particularly in the solid state, might be modified by external environments. Indeed, although the suppression we describe is deleterious for dimer photocell currents, one could imagine that the ability to dynamically suppress transport by a real-time change in an external environmental property (temperature, pressure, etc.) might be useful for rectification or gating a signal, perhaps even allowing for sensing functionalities.

Interestingly, it has recently become possible to program and implement Hamiltonian models of light harvesting in `quantum simulators', as recently achieved with superconducting qubits \cite{anton}, ion traps \cite{PRX}, and a NMR system \cite{neill}. In the case of superconducting qubits, dephasing noise of arbitrary spectral properties can be applied to each individual `chromophore' and their quantum coupling to photons can also be tuned in strength and frequency, allowing, in principle, a precise experimental test of non-additive environmental effects in both Markovian and non-Markovian photosynthetic energy transport. Further avenues for future study include the influence of coexisting underdamped and overdamped structured environments \cite{Iles_Smith_2015,Santamore2013,elinor1,Castro2014,killoran2015, Stones2016,Strasberg_2016, elinor2, Chen2016,Newman_2016, Qin2017} in dimer photocells, and how off-resonant dimers may alter the effects we discuss herein \cite{amir2015}.

\acknowledgements
We would like to thank Prof. Matthias Troyer for creating the opportunity for this master's project, and Ahsan Nazir, Henry Maguire, and Mauro Cirio for helpful discussions and feedback on this work. M.W. has been fully supported by the RIKEN IPA program. N.L. and F.N. acknowledge
support from the RIKEN-AIST Joint Research Fund and the Sir John Templeton Foundation. FN is partially supported by the
MURI Center for Dynamic Magneto-Optics via the AFOSR Award No. FA9550-14-1-0040, the Army Research Office (ARO) under grant number 73315PH, the AOARD grant No.~ FA2386-18-1-4045, the IMPACT program of JST,
CREST Grant No. JPMJCR1676,
JSPS-RFBR Grant No. 17-52-50023.

\section{Appendix}

In our results section, we compare results from the reaction coordinate master equation, \eqr{RCME}, with two Markovian master equations which treat the interaction with the phonon environment under a standard Born-Markov assumption, i.e., as a rate $\gamma_x$ driving transitions between the bright and dark states of the system.  The secular version of this master equation is explicitly defined, using the same notation as in the main text, as,
\begin{equation}
\begin{split}
\dot{\rho}(t)&=-i[H_S, \rho(t)] \\
&+ \mathcal{L}_{x_1 b}[\rho(t)]+ \mathcal{L}_{b x_1}[\rho(t)]\\
&+  \mathcal{L}_{x_2 x_1}[\rho(t)] +  \mathcal{L}_{x_1 x_2}[\rho(t)]  \\
&+\mathcal{L}_{\alpha x_2}[\rho(t)]+\mathcal{L}_{x_2 \alpha}[\rho(t)] \\
&+ \mathcal{L}_{\beta b}[\rho(t)]+ \mathcal{L}_{b\beta}[\rho(t)] \\
&+ \mathcal{L}_{\alpha\beta}[\rho(t)]+ \mathcal{L}_{\alpha b}[\rho(t)],
\end{split}
\label{SME}
\end{equation}

Now, all transitions are given by Lindblad generators $\mathcal{L}_{AB}= \left[C_{AB}\rho(t)C_{AB}^{\dagger}- \frac{1}{2} \left\{C_{AB}^{\dagger}C_{AB},\rho(t)\right\}\right]$.
In addition to the parameters $C_{AB}$ defined in \eqr{c_ops}, we also have
\begin{equation}
\begin{split}
C_{x_1 b} &= \sqrt{2\gamma_h n_h}\ket{x_1}\bra{b}\\
C_{b x_1} &= \sqrt{2\gamma_h(1+n_h})\ket{b}\bra{x_1}\\
C_{x_2 x_1} &= \sqrt{\gamma_x(n_x+1)}\ket{x_2}\bra{x_1}\\
C_{x_1 x_2} &= \sqrt{\gamma_x n_x}\ket{x_1}\bra{x_2}\\
C_{\alpha x_2} &= \sqrt{2\gamma_c(n_c+1)}\ket{\alpha}\bra{x_2}\\
C_{x_2 \alpha} &= \sqrt{2\gamma_c n_c}\ket{x_2}\bra{\alpha}.\\
\end{split}
\end{equation}
The factors of $2$ in the photon baths and the transitions to the state $\ket{\alpha}$ arise from the collective dipole coupling, such that the matrix elements for these transitions are enhanced over those in the bare $\ket{a_1}$ and $\ket{a_2}$ basis.

Even in this case of a Markovian treatment of the phonon environment, the above equation is incomplete when the rates approach the coupling between the dimers, $2J$.  In this case, one must treat transitions involving the excited states of the dimer with a non-Secular Born-Markov master equation.
The equation of motion then becomes,
\begin{equation}
\begin{split}
\dot{\rho}(t)&=-i[H_S, \rho(t)] + \mathcal{M}_{s}[\rho(t)]+ \mathcal{M}_{Q_h}[\rho(t)]+ \mathcal{M}_{Q_c^{(\alpha)}}[\rho(t)] \\
&+ \mathcal{L}_{\beta b}[\rho(t)]+ \mathcal{L}_{b\beta}[\rho(t)] + \mathcal{L}_{\alpha\beta}[\rho(t)]+ \mathcal{L}_{\alpha b}[\rho(t)],
\end{split}
\label{NSME}
\end{equation}
Again, the transitions not involving the excited dimer states are given by secular Born-Markov Lindblad generators.  The transitions involving the hot photon bath, and those involving the electron transfer process to $\ket{\alpha}$, are given by the non-secular generator \eqr{MQ}, (although now the eigenstates of the system do not include the RC mode).
The Markovian phonon transitions are also given by \eqr{MQ} but with operator  $s = (\ket{a_1}\bra{a_1}-\ket{a_2}\bra{a_2})$ and spectral density $J_{s}(\omega) = J_0(\omega)$, as per \eqr{eq:SD_0}.

\bibliography{Masterarbeit}

\begin{thebibliography}{65}%
\makeatletter
\providecommand \@ifxundefined [1]{%
 \@ifx{#1\undefined}
}%
\providecommand \@ifnum [1]{%
 \ifnum #1\expandafter \@firstoftwo
 \else \expandafter \@secondoftwo
 \fi
}%
\providecommand \@ifx [1]{%
 \ifx #1\expandafter \@firstoftwo
 \else \expandafter \@secondoftwo
 \fi
}%
\providecommand \natexlab [1]{#1}%
\providecommand \enquote  [1]{``#1''}%
\providecommand \bibnamefont  [1]{#1}%
\providecommand \bibfnamefont [1]{#1}%
\providecommand \citenamefont [1]{#1}%
\providecommand \href@noop [0]{\@secondoftwo}%
\providecommand \href [0]{\begingroup \@sanitize@url \@href}%
\providecommand \@href[1]{\@@startlink{#1}\@@href}%
\providecommand \@@href[1]{\endgroup#1\@@endlink}%
\providecommand \@sanitize@url [0]{\catcode `\\12\catcode `\$12\catcode
  `\&12\catcode `\#12\catcode `\^12\catcode `\_12\catcode `\%12\relax}%
\providecommand \@@startlink[1]{}%
\providecommand \@@endlink[0]{}%
\providecommand \url  [0]{\begingroup\@sanitize@url \@url }%
\providecommand \@url [1]{\endgroup\@href {#1}{\urlprefix }}%
\providecommand \urlprefix  [0]{URL }%
\providecommand \Eprint [0]{\href }%
\providecommand \doibase [0]{http://dx.doi.org/}%
\providecommand \selectlanguage [0]{\@gobble}%
\providecommand \bibinfo  [0]{\@secondoftwo}%
\providecommand \bibfield  [0]{\@secondoftwo}%
\providecommand \translation [1]{[#1]}%
\providecommand \BibitemOpen [0]{}%
\providecommand \bibitemStop [0]{}%
\providecommand \bibitemNoStop [0]{.\EOS\space}%
\providecommand \EOS [0]{\spacefactor3000\relax}%
\providecommand \BibitemShut  [1]{\csname bibitem#1\endcsname}%
\let\auto@bib@innerbib\@empty
\bibitem [{\citenamefont {Cheng}\ and\ \citenamefont
  {Fleming}(2009)}]{chengreview}%
  \BibitemOpen
  \bibfield  {author} {\bibinfo {author} {\bibfnamefont {Y.-C.}\ \bibnamefont
  {Cheng}}\ and\ \bibinfo {author} {\bibfnamefont {G.~R.}\ \bibnamefont
  {Fleming}},\ }\bibfield  {title} {\enquote {\bibinfo {title} {Dynamics of
  light harvesting in photosynthesis},}\ }\href {\doibase
  10.1146/annurev.physchem.040808.090259} {\bibfield  {journal} {\bibinfo
  {journal} {Annual Review of Physical Chemistry}\ }\textbf {\bibinfo {volume}
  {60}},\ \bibinfo {pages} {241--262} (\bibinfo {year} {2009})},\ \bibinfo
  {note} {pMID: 18999996},\ \Eprint
  {http://arxiv.org/abs/https://doi.org/10.1146/annurev.physchem.040808.090259}
  {https://doi.org/10.1146/annurev.physchem.040808.090259} \BibitemShut
  {NoStop}%
\bibitem [{\citenamefont {Ishizaki}\ and\ \citenamefont
  {Fleming}(2012)}]{ishizakireview}%
  \BibitemOpen
  \bibfield  {author} {\bibinfo {author} {\bibfnamefont {A.}~\bibnamefont
  {Ishizaki}}\ and\ \bibinfo {author} {\bibfnamefont {G.~R.}\ \bibnamefont
  {Fleming}},\ }\bibfield  {title} {\enquote {\bibinfo {title} {Quantum
  coherence in photosynthetic light harvesting},}\ }\href {\doibase
  10.1146/annurev-conmatphys-020911-125126} {\bibfield  {journal} {\bibinfo
  {journal} {Annual Review of Condensed Matter Physics}\ }\textbf {\bibinfo
  {volume} {3}},\ \bibinfo {pages} {333--361} (\bibinfo {year} {2012})},\
  \Eprint
  {http://arxiv.org/abs/https://doi.org/10.1146/annurev-conmatphys-020911-125126}
  {https://doi.org/10.1146/annurev-conmatphys-020911-125126} \BibitemShut
  {NoStop}%
\bibitem [{\citenamefont {Lambert}\ \emph {et~al.}(2013)\citenamefont
  {Lambert}, \citenamefont {Chen}, \citenamefont {Cheng}, \citenamefont {Li},
  \citenamefont {Chen},\ and\ \citenamefont {Nori}}]{Lambert_2013}%
  \BibitemOpen
  \bibfield  {author} {\bibinfo {author} {\bibfnamefont {N.}~\bibnamefont
  {Lambert}}, \bibinfo {author} {\bibfnamefont {Y.~N.}\ \bibnamefont {Chen}},
  \bibinfo {author} {\bibfnamefont {Y.~C.}\ \bibnamefont {Cheng}}, \bibinfo
  {author} {\bibfnamefont {C.~M.}\ \bibnamefont {Li}}, \bibinfo {author}
  {\bibfnamefont {G.~Y.}\ \bibnamefont {Chen}}, \ and\ \bibinfo {author}
  {\bibfnamefont {F.}~\bibnamefont {Nori}},\ }\bibfield  {title} {\enquote
  {\bibinfo {title} {Quantum biology},}\ }\href {\doibase 10.1038/nphys2474}
  {\bibfield  {journal} {\bibinfo  {journal} {Nat. Phys.}\ }\textbf {\bibinfo
  {volume} {9}},\ \bibinfo {pages} {10--18} (\bibinfo {year}
  {2013})}\BibitemShut {NoStop}%
\bibitem [{\citenamefont {Scholes}\ \emph {et~al.}(2017)\citenamefont
  {Scholes}, \citenamefont {Fleming}, \citenamefont {Chen}, \citenamefont
  {Aspuru-Guzik}, \citenamefont {Buchleitner}, \citenamefont {Coker},
  \citenamefont {Engel}, \citenamefont {van Grondelle}, \citenamefont
  {Ishizaki}, \citenamefont {Jonas}, \citenamefont {Lundeen}, \citenamefont
  {McCusker}, \citenamefont {Mukamel}, \citenamefont {Ogilvie}, \citenamefont
  {Olaya-Castro}, \citenamefont {Ratner}, \citenamefont {Spano}, \citenamefont
  {Whaley},\ and\ \citenamefont {Zhu}}]{Scholes2017}%
  \BibitemOpen
  \bibfield  {author} {\bibinfo {author} {\bibfnamefont {G.~D.}\ \bibnamefont
  {Scholes}}, \bibinfo {author} {\bibfnamefont {G.~R.}\ \bibnamefont
  {Fleming}}, \bibinfo {author} {\bibfnamefont {L.~X.}\ \bibnamefont {Chen}},
  \bibinfo {author} {\bibfnamefont {A.}~\bibnamefont {Aspuru-Guzik}}, \bibinfo
  {author} {\bibfnamefont {A.}~\bibnamefont {Buchleitner}}, \bibinfo {author}
  {\bibfnamefont {D.~F.}\ \bibnamefont {Coker}}, \bibinfo {author}
  {\bibfnamefont {G.~S.}\ \bibnamefont {Engel}}, \bibinfo {author}
  {\bibfnamefont {R.}~\bibnamefont {van Grondelle}}, \bibinfo {author}
  {\bibfnamefont {A.}~\bibnamefont {Ishizaki}}, \bibinfo {author}
  {\bibfnamefont {D.~M.}\ \bibnamefont {Jonas}}, \bibinfo {author}
  {\bibfnamefont {J.~S.}\ \bibnamefont {Lundeen}}, \bibinfo {author}
  {\bibfnamefont {J.~K.}\ \bibnamefont {McCusker}}, \bibinfo {author}
  {\bibfnamefont {S.}~\bibnamefont {Mukamel}}, \bibinfo {author} {\bibfnamefont
  {J.~P.}\ \bibnamefont {Ogilvie}}, \bibinfo {author} {\bibfnamefont
  {A.}~\bibnamefont {Olaya-Castro}}, \bibinfo {author} {\bibfnamefont {M.~A.}\
  \bibnamefont {Ratner}}, \bibinfo {author} {\bibfnamefont {F.~C.}\
  \bibnamefont {Spano}}, \bibinfo {author} {\bibfnamefont {K.~B.}\ \bibnamefont
  {Whaley}}, \ and\ \bibinfo {author} {\bibfnamefont {X.}~\bibnamefont {Zhu}},\
  }\bibfield  {title} {\enquote {\bibinfo {title} {{Using coherence to enhance
  function in chemical and biophysical systems}},}\ }\href {\doibase
  10.1038/nature21425} {\bibfield  {journal} {\bibinfo  {journal} {Nature}\
  }\textbf {\bibinfo {volume} {543}},\ \bibinfo {pages} {647--656} (\bibinfo
  {year} {2017})}\BibitemShut {NoStop}%
\bibitem [{\citenamefont {Engel}\ \emph {et~al.}(2007)\citenamefont {Engel},
  \citenamefont {Calhoun}, \citenamefont {Read}, \citenamefont {Ahn},
  \citenamefont {Man\v{c}al}, \citenamefont {Cheng}, \citenamefont
  {Blankenship},\ and\ \citenamefont {Fleming}}]{Engel2007}%
  \BibitemOpen
  \bibfield  {author} {\bibinfo {author} {\bibfnamefont {G.~S.}\ \bibnamefont
  {Engel}}, \bibinfo {author} {\bibfnamefont {T.~R.}\ \bibnamefont {Calhoun}},
  \bibinfo {author} {\bibfnamefont {E.~L.}\ \bibnamefont {Read}}, \bibinfo
  {author} {\bibfnamefont {T.-K.}\ \bibnamefont {Ahn}}, \bibinfo {author}
  {\bibfnamefont {T.}~\bibnamefont {Man\v{c}al}}, \bibinfo {author}
  {\bibfnamefont {Y.-C.}\ \bibnamefont {Cheng}}, \bibinfo {author}
  {\bibfnamefont {R.~E.}\ \bibnamefont {Blankenship}}, \ and\ \bibinfo {author}
  {\bibfnamefont {G.~R.}\ \bibnamefont {Fleming}},\ }\bibfield  {title}
  {\enquote {\bibinfo {title} {Evidence for wavelike energy transfer through
  quantum coherence in photosynthetic systems},}\ }\href {\doibase
  10.1038/nature05678} {\bibfield  {journal} {\bibinfo  {journal} {Nature}\ }
  (\bibinfo {year} {2007}),\ 10.1038/nature05678}\BibitemShut {NoStop}%
\bibitem [{\citenamefont {Ishizaki}\ and\ \citenamefont
  {Fleming}(2009)}]{Akihito09}%
  \BibitemOpen
  \bibfield  {author} {\bibinfo {author} {\bibfnamefont {A.}~\bibnamefont
  {Ishizaki}}\ and\ \bibinfo {author} {\bibfnamefont {G.~R.}\ \bibnamefont
  {Fleming}},\ }\bibfield  {title} {\enquote {\bibinfo {title} {{Theoretical
  examination of quantum coherence in a photosynthetic system at physiological
  temperature}},}\ }\href {\doibase 10.1073/pnas.0908989106} {\bibfield
  {journal} {\bibinfo  {journal} {Proc. Natl. Acad. Sci. USA}\ }\textbf
  {\bibinfo {volume} {106}},\ \bibinfo {pages} {17255--17260} (\bibinfo {year}
  {2009})}\BibitemShut {NoStop}%
\bibitem [{\citenamefont {Collini}\ \emph {et~al.}(2010)\citenamefont
  {Collini}, \citenamefont {Wong}, \citenamefont {Wilk}, \citenamefont {Curmi},
  \citenamefont {Brumer},\ and\ \citenamefont {Scholes}}]{Collini2010}%
  \BibitemOpen
  \bibfield  {author} {\bibinfo {author} {\bibfnamefont {E.}~\bibnamefont
  {Collini}}, \bibinfo {author} {\bibfnamefont {C.~Y.}\ \bibnamefont {Wong}},
  \bibinfo {author} {\bibfnamefont {K.~E.}\ \bibnamefont {Wilk}}, \bibinfo
  {author} {\bibfnamefont {P.~M.~G.}\ \bibnamefont {Curmi}}, \bibinfo {author}
  {\bibfnamefont {P.}~\bibnamefont {Brumer}}, \ and\ \bibinfo {author}
  {\bibfnamefont {G.~D.}\ \bibnamefont {Scholes}},\ }\bibfield  {title}
  {\enquote {\bibinfo {title} {Coherently wired light-harvesting in
  photosynthetic marine algae at ambient temperature},}\ }\href {\doibase
  10.1038/nature08811} {\bibfield  {journal} {\bibinfo  {journal} {Nature}\ }
  (\bibinfo {year} {2010}),\ 10.1038/nature08811}\BibitemShut {NoStop}%
\bibitem [{\citenamefont {Plenio}\ and\ \citenamefont
  {Huelga}(2008)}]{Plenio08}%
  \BibitemOpen
  \bibfield  {author} {\bibinfo {author} {\bibfnamefont {M.~B.}\ \bibnamefont
  {Plenio}}\ and\ \bibinfo {author} {\bibfnamefont {S.~F.}\ \bibnamefont
  {Huelga}},\ }\bibfield  {title} {\enquote {\bibinfo {title}
  {{Dephasing-assisted transport: quantum networks and biomolecules}},}\ }\href
  {\doibase 10.1088/1367-2630/10/11/113019} {\bibfield  {journal} {\bibinfo
  {journal} {New J. Phys.}\ }\textbf {\bibinfo {volume} {10}},\ \bibinfo
  {pages} {113019} (\bibinfo {year} {2008})}\BibitemShut {NoStop}%
\bibitem [{\citenamefont {{P. Rebentrost}}\ \emph {et~al.}(2009)\citenamefont
  {{P. Rebentrost}}, \citenamefont {{M. Mohseni}}, \citenamefont {{I. Kassal}},
  \citenamefont {{S. Lloyd}},\ and\ \citenamefont
  {Aspuru-Guzik}}]{RebentrostEtAlNJP2009}%
  \BibitemOpen
  \bibfield  {author} {\bibinfo {author} {\bibnamefont {{P. Rebentrost}}},
  \bibinfo {author} {\bibnamefont {{M. Mohseni}}}, \bibinfo {author}
  {\bibnamefont {{I. Kassal}}}, \bibinfo {author} {\bibnamefont {{S. Lloyd}}},
  \ and\ \bibinfo {author} {\bibfnamefont {A.}~\bibnamefont {Aspuru-Guzik}},\
  }\bibfield  {title} {\enquote {\bibinfo {title} {{Environment-assisted
  quantum transport}},}\ }\href {\doibase 10.1088/1367-2630/11/3/033003}
  {\bibfield  {journal} {\bibinfo  {journal} {New J. Phys.}\ }\textbf {\bibinfo
  {volume} {11}},\ \bibinfo {pages} {33003} (\bibinfo {year}
  {2009})}\BibitemShut {NoStop}%
\bibitem [{\citenamefont {Caruso}\ \emph {et~al.}(2009)\citenamefont {Caruso},
  \citenamefont {Chin}, \citenamefont {Datta}, \citenamefont {Huelga},\ and\
  \citenamefont {Plenio}}]{Caruso09}%
  \BibitemOpen
  \bibfield  {author} {\bibinfo {author} {\bibfnamefont {F.}~\bibnamefont
  {Caruso}}, \bibinfo {author} {\bibfnamefont {A.~W.}\ \bibnamefont {Chin}},
  \bibinfo {author} {\bibfnamefont {A.}~\bibnamefont {Datta}}, \bibinfo
  {author} {\bibfnamefont {S.~F.}\ \bibnamefont {Huelga}}, \ and\ \bibinfo
  {author} {\bibfnamefont {M.~B.}\ \bibnamefont {Plenio}},\ }\bibfield  {title}
  {\enquote {\bibinfo {title} {{Highly efficient energy excitation transer in
  light-harvesting complexes: The fundamental role of noise-assisted
  transport}},}\ }\href {\doibase 10.1063/1.3223548} {\bibfield  {journal}
  {\bibinfo  {journal} {J. Chem. Phys.}\ }\textbf {\bibinfo {volume} {131}},\
  \bibinfo {pages} {105106} (\bibinfo {year} {2009})}\BibitemShut {NoStop}%
\bibitem [{\citenamefont {Chin}\ \emph {et~al.}(2010)\citenamefont {Chin},
  \citenamefont {Datta}, \citenamefont {Caruso}, \citenamefont {Plenio},\ and\
  \citenamefont {Huelga}}]{Plenio10}%
  \BibitemOpen
  \bibfield  {author} {\bibinfo {author} {\bibfnamefont {A.~W.}\ \bibnamefont
  {Chin}}, \bibinfo {author} {\bibfnamefont {A.}~\bibnamefont {Datta}},
  \bibinfo {author} {\bibfnamefont {F.}~\bibnamefont {Caruso}}, \bibinfo
  {author} {\bibfnamefont {M.~B.}\ \bibnamefont {Plenio}}, \ and\ \bibinfo
  {author} {\bibfnamefont {S.~F.}\ \bibnamefont {Huelga}},\ }\bibfield  {title}
  {\enquote {\bibinfo {title} {{Noise-assisted energy transfer in quantum
  networks and light-harvesting complexes}},}\ }\href {\doibase
  10.1088/1367-2630/12/6/065002} {\bibfield  {journal} {\bibinfo  {journal}
  {New J. Phys.}\ }\textbf {\bibinfo {volume} {12}},\ \bibinfo {pages} {065002}
  (\bibinfo {year} {2010})}\BibitemShut {NoStop}%
\bibitem [{\citenamefont {Caycedo-Soler}\ \emph {et~al.}(2010)\citenamefont
  {Caycedo-Soler}, \citenamefont {Rodriguez}, \citenamefont {Quiroga},\ and\
  \citenamefont {Johnson}}]{bleaching}%
  \BibitemOpen
  \bibfield  {author} {\bibinfo {author} {\bibfnamefont {F.}~\bibnamefont
  {Caycedo-Soler}}, \bibinfo {author} {\bibfnamefont {F.~J.}\ \bibnamefont
  {Rodriguez}}, \bibinfo {author} {\bibfnamefont {L.}~\bibnamefont {Quiroga}},
  \ and\ \bibinfo {author} {\bibfnamefont {N.~F.}\ \bibnamefont {Johnson}},\
  }\bibfield  {title} {\enquote {\bibinfo {title} {{Light-harvesting mechanism
  of bacteria exploits a critical interplay between the dynamics of transport
  and trapping}},}\ }\href {\doibase 10.1103/physrevlett.104.158302} {\bibfield
   {journal} {\bibinfo  {journal} {Phys. Rev. Lett.}\ }\textbf {\bibinfo
  {volume} {104}},\ \bibinfo {pages} {158302} (\bibinfo {year}
  {2010})}\BibitemShut {NoStop}%
\bibitem [{\citenamefont {Chin}\ \emph
  {et~al.}(2013{\natexlab{a}})\citenamefont {Chin}, \citenamefont {Prior},
  \citenamefont {Rosenbach}, \citenamefont {Caycedo-Soler}, \citenamefont
  {Huelga},\ and\ \citenamefont {Plenio}}]{Chin2013}%
  \BibitemOpen
  \bibfield  {author} {\bibinfo {author} {\bibfnamefont {A.~W.}\ \bibnamefont
  {Chin}}, \bibinfo {author} {\bibfnamefont {J.}~\bibnamefont {Prior}},
  \bibinfo {author} {\bibfnamefont {R.}~\bibnamefont {Rosenbach}}, \bibinfo
  {author} {\bibfnamefont {F.}~\bibnamefont {Caycedo-Soler}}, \bibinfo {author}
  {\bibfnamefont {S.~F.}\ \bibnamefont {Huelga}}, \ and\ \bibinfo {author}
  {\bibfnamefont {M.~B.}\ \bibnamefont {Plenio}},\ }\bibfield  {title}
  {\enquote {\bibinfo {title} {The role of non-equilibrium vibrational
  structures in electronic coherence and recoherence in pigment-protein
  complexes},}\ }\href {http://dx.doi.org/10.1038/nphys2515} {\bibfield
  {journal} {\bibinfo  {journal} {Nat Phys}\ }\textbf {\bibinfo {volume} {9}},\
  \bibinfo {pages} {113--118} (\bibinfo {year}
  {2013}{\natexlab{a}})}\BibitemShut {NoStop}%
\bibitem [{\citenamefont {Li}\ \emph {et~al.}(2012)\citenamefont {Li},
  \citenamefont {Lambert}, \citenamefont {Chen}, \citenamefont {Chen},\ and\
  \citenamefont {Nori}}]{Li2012}%
  \BibitemOpen
  \bibfield  {author} {\bibinfo {author} {\bibfnamefont {C.-M.}\ \bibnamefont
  {Li}}, \bibinfo {author} {\bibfnamefont {N.}~\bibnamefont {Lambert}},
  \bibinfo {author} {\bibfnamefont {Y.-N.}\ \bibnamefont {Chen}}, \bibinfo
  {author} {\bibfnamefont {G.-Y.}\ \bibnamefont {Chen}}, \ and\ \bibinfo
  {author} {\bibfnamefont {F.}~\bibnamefont {Nori}},\ }\bibfield  {title}
  {\enquote {\bibinfo {title} {{Witnessing quantum coherence: from solid-state
  to biological systems}},}\ }\href {\doibase 10.1038/srep00885} {\bibfield
  {journal} {\bibinfo  {journal} {Scientific Reports}\ }\textbf {\bibinfo
  {volume} {2}},\ \bibinfo {pages} {885} (\bibinfo {year} {2012})},\ \Eprint
  {http://arxiv.org/abs/1212.0194} {arXiv:1212.0194} \BibitemShut {NoStop}%
\bibitem [{\citenamefont {Mourokh}\ and\ \citenamefont
  {Nori}(2015)}]{PhysRevE.92.052720}%
  \BibitemOpen
  \bibfield  {author} {\bibinfo {author} {\bibfnamefont {L.~G.}\ \bibnamefont
  {Mourokh}}\ and\ \bibinfo {author} {\bibfnamefont {F.}~\bibnamefont {Nori}},\
  }\bibfield  {title} {\enquote {\bibinfo {title} {Energy transfer efficiency
  in the chromophore network strongly coupled to a vibrational mode},}\ }\href
  {\doibase 10.1103/PhysRevE.92.052720} {\bibfield  {journal} {\bibinfo
  {journal} {Phys. Rev. E}\ }\textbf {\bibinfo {volume} {92}},\ \bibinfo
  {pages} {052720} (\bibinfo {year} {2015})}\BibitemShut {NoStop}%
\bibitem [{\citenamefont {Chin}, \citenamefont {Huelga},\ and\ \citenamefont
  {Plenio}(2012)}]{Chin2012}%
  \BibitemOpen
  \bibfield  {author} {\bibinfo {author} {\bibfnamefont {A.~W.}\ \bibnamefont
  {Chin}}, \bibinfo {author} {\bibfnamefont {S.~F.}\ \bibnamefont {Huelga}}, \
  and\ \bibinfo {author} {\bibfnamefont {M.~B.}\ \bibnamefont {Plenio}},\
  }\bibfield  {title} {\enquote {\bibinfo {title} {Coherence and decoherence in
  biological systems: Principles of noise assisted transport and the origin of
  long-lived coherences},}\ }\href {\doibase 10.1098/rsta.2011.0224} {\bibfield
   {journal} {\bibinfo  {journal} {Phil. Trans. R. Soc. A}\ }\textbf {\bibinfo
  {volume} {370}},\ \bibinfo {pages} {3638} (\bibinfo {year} {2012})},\ \Eprint
  {http://arxiv.org/abs/1203.5072v1} {1203.5072v1} \BibitemShut {NoStop}%
\bibitem [{\citenamefont {Huelga}\ and\ \citenamefont
  {Plenio}(2013)}]{Huelga2013}%
  \BibitemOpen
  \bibfield  {author} {\bibinfo {author} {\bibfnamefont {S.~F.}\ \bibnamefont
  {Huelga}}\ and\ \bibinfo {author} {\bibfnamefont {M.~B.}\ \bibnamefont
  {Plenio}},\ }\bibfield  {title} {\enquote {\bibinfo {title} {Vibrations,
  quanta and biology},}\ }\href {\doibase 10.1080/00405000.2013.829687}
  {\bibfield  {journal} {\bibinfo  {journal} {Contemp. Phys.}\ }\textbf
  {\bibinfo {volume} {54}},\ \bibinfo {pages} {181} (\bibinfo {year} {2013})},\
  \Eprint {http://arxiv.org/abs/1307.3530v1} {1307.3530v1} \BibitemShut
  {NoStop}%
\bibitem [{\citenamefont {Romero}\ \emph {et~al.}(2014)\citenamefont {Romero},
  \citenamefont {Augulis}, \citenamefont {Novoderezhkin}, \citenamefont
  {Ferretti}, \citenamefont {Thieme}, \citenamefont {Zigmantas},\ and\
  \citenamefont {Van~Grondelle}}]{romero2014quantum}%
  \BibitemOpen
  \bibfield  {author} {\bibinfo {author} {\bibfnamefont {E.}~\bibnamefont
  {Romero}}, \bibinfo {author} {\bibfnamefont {R.}~\bibnamefont {Augulis}},
  \bibinfo {author} {\bibfnamefont {V.~I.}\ \bibnamefont {Novoderezhkin}},
  \bibinfo {author} {\bibfnamefont {M.}~\bibnamefont {Ferretti}}, \bibinfo
  {author} {\bibfnamefont {J.}~\bibnamefont {Thieme}}, \bibinfo {author}
  {\bibfnamefont {D.}~\bibnamefont {Zigmantas}}, \ and\ \bibinfo {author}
  {\bibfnamefont {R.}~\bibnamefont {Van~Grondelle}},\ }\bibfield  {title}
  {\enquote {\bibinfo {title} {Quantum coherence in photosynthesis for
  efficient solar-energy conversion},}\ }\href@noop {} {\bibfield  {journal}
  {\bibinfo  {journal} {Nature physics}\ }\textbf {\bibinfo {volume} {10}},\
  \bibinfo {pages} {676} (\bibinfo {year} {2014})}\BibitemShut {NoStop}%
\bibitem [{\citenamefont {Fuller}\ \emph {et~al.}(2014)\citenamefont {Fuller},
  \citenamefont {Pan}, \citenamefont {Gelzinis}, \citenamefont {Butkus},
  \citenamefont {Senlik}, \citenamefont {Wilcox}, \citenamefont {Yocum},
  \citenamefont {Valkunas}, \citenamefont {Abramavicius},\ and\ \citenamefont
  {Ogilvie}}]{fuller2014vibronic}%
  \BibitemOpen
  \bibfield  {author} {\bibinfo {author} {\bibfnamefont {F.~D.}\ \bibnamefont
  {Fuller}}, \bibinfo {author} {\bibfnamefont {J.}~\bibnamefont {Pan}},
  \bibinfo {author} {\bibfnamefont {A.}~\bibnamefont {Gelzinis}}, \bibinfo
  {author} {\bibfnamefont {V.}~\bibnamefont {Butkus}}, \bibinfo {author}
  {\bibfnamefont {S.~S.}\ \bibnamefont {Senlik}}, \bibinfo {author}
  {\bibfnamefont {D.~E.}\ \bibnamefont {Wilcox}}, \bibinfo {author}
  {\bibfnamefont {C.~F.}\ \bibnamefont {Yocum}}, \bibinfo {author}
  {\bibfnamefont {L.}~\bibnamefont {Valkunas}}, \bibinfo {author}
  {\bibfnamefont {D.}~\bibnamefont {Abramavicius}}, \ and\ \bibinfo {author}
  {\bibfnamefont {J.~P.}\ \bibnamefont {Ogilvie}},\ }\bibfield  {title}
  {\enquote {\bibinfo {title} {Vibronic coherence in oxygenic
  photosynthesis},}\ }\href@noop {} {\bibfield  {journal} {\bibinfo  {journal}
  {Nature chemistry}\ }\textbf {\bibinfo {volume} {6}},\ \bibinfo {pages} {706}
  (\bibinfo {year} {2014})}\BibitemShut {NoStop}%
\bibitem [{\citenamefont {Dorfman}\ \emph {et~al.}(2013)\citenamefont
  {Dorfman}, \citenamefont {Voronine}, \citenamefont {Mukamel},\ and\
  \citenamefont {Scully}}]{Dorfman_2013}%
  \BibitemOpen
  \bibfield  {author} {\bibinfo {author} {\bibfnamefont {K.~E.}\ \bibnamefont
  {Dorfman}}, \bibinfo {author} {\bibfnamefont {D.~V.}\ \bibnamefont
  {Voronine}}, \bibinfo {author} {\bibfnamefont {S.}~\bibnamefont {Mukamel}}, \
  and\ \bibinfo {author} {\bibfnamefont {M.~O.}\ \bibnamefont {Scully}},\
  }\bibfield  {title} {\enquote {\bibinfo {title} {Photosynthetic reaction
  center as a quantum heat engine},}\ }\href {\doibase 10.1073/pnas.1212666110}
  {\bibfield  {journal} {\bibinfo  {journal} {Proceedings of the National
  Academy of Sciences}\ }\textbf {\bibinfo {volume} {110}},\ \bibinfo {pages}
  {2746--2751} (\bibinfo {year} {2013})}\BibitemShut {NoStop}%
\bibitem [{\citenamefont {Creatore}\ \emph {et~al.}(2013)\citenamefont
  {Creatore}, \citenamefont {Parker}, \citenamefont {Emmott},\ and\
  \citenamefont {Chin}}]{Creatore_2013}%
  \BibitemOpen
  \bibfield  {author} {\bibinfo {author} {\bibfnamefont {C.}~\bibnamefont
  {Creatore}}, \bibinfo {author} {\bibfnamefont {M.~A.}\ \bibnamefont
  {Parker}}, \bibinfo {author} {\bibfnamefont {S.}~\bibnamefont {Emmott}}, \
  and\ \bibinfo {author} {\bibfnamefont {A.~W.}\ \bibnamefont {Chin}},\
  }\bibfield  {title} {\enquote {\bibinfo {title} {Efficient biologically
  inspired photocell enhanced by delocalized quantum states},}\ }\href
  {\doibase 10.1103/physrevlett.111.253601} {\bibfield  {journal} {\bibinfo
  {journal} {Physical Review Letters}\ }\textbf {\bibinfo {volume} {111}}
  (\bibinfo {year} {2013}),\ 10.1103/physrevlett.111.253601}\BibitemShut
  {NoStop}%
\bibitem [{\citenamefont {Zhang}\ \emph {et~al.}(2015)\citenamefont {Zhang},
  \citenamefont {Oh}, \citenamefont {Alharbi}, \citenamefont {Engel},\ and\
  \citenamefont {Kais}}]{Zhang_2015}%
  \BibitemOpen
  \bibfield  {author} {\bibinfo {author} {\bibfnamefont {Y.}~\bibnamefont
  {Zhang}}, \bibinfo {author} {\bibfnamefont {S.}~\bibnamefont {Oh}}, \bibinfo
  {author} {\bibfnamefont {F.~H.}\ \bibnamefont {Alharbi}}, \bibinfo {author}
  {\bibfnamefont {G.~S.}\ \bibnamefont {Engel}}, \ and\ \bibinfo {author}
  {\bibfnamefont {S.}~\bibnamefont {Kais}},\ }\bibfield  {title} {\enquote
  {\bibinfo {title} {Delocalized quantum states enhance photocell
  efficiency},}\ }\href {\doibase 10.1039/c4cp05310a} {\bibfield  {journal}
  {\bibinfo  {journal} {Phys. Chem. Chem. Phys.}\ }\textbf {\bibinfo {volume}
  {17}},\ \bibinfo {pages} {5743--5750} (\bibinfo {year} {2015})}\BibitemShut
  {NoStop}%
\bibitem [{\citenamefont {Shockley}\ and\ \citenamefont
  {Queisser}(1961)}]{Shockley_1961}%
  \BibitemOpen
  \bibfield  {author} {\bibinfo {author} {\bibfnamefont {W.}~\bibnamefont
  {Shockley}}\ and\ \bibinfo {author} {\bibfnamefont {H.~J.}\ \bibnamefont
  {Queisser}},\ }\bibfield  {title} {\enquote {\bibinfo {title} {Detailed
  balance limit of efficiency of p-n junction solar cells},}\ }\href {\doibase
  10.1063/1.1736034} {\bibfield  {journal} {\bibinfo  {journal} {Journal of
  Applied Physics}\ }\textbf {\bibinfo {volume} {32}},\ \bibinfo {pages}
  {510--519} (\bibinfo {year} {1961})}\BibitemShut {NoStop}%
\bibitem [{\citenamefont {Santamore}, \citenamefont {Lambert},\ and\
  \citenamefont {Nori}(2013)}]{Santamore2013}%
  \BibitemOpen
  \bibfield  {author} {\bibinfo {author} {\bibfnamefont {D.~H.}\ \bibnamefont
  {Santamore}}, \bibinfo {author} {\bibfnamefont {N.}~\bibnamefont {Lambert}},
  \ and\ \bibinfo {author} {\bibfnamefont {F.}~\bibnamefont {Nori}},\
  }\bibfield  {title} {\enquote {\bibinfo {title} {{Vibrationally mediated
  transport in molecular transistors}},}\ }\href {\doibase
  10.1103/PhysRevB.87.075422} {\bibfield  {journal} {\bibinfo  {journal} {Phys.
  Rev. B}\ }\textbf {\bibinfo {volume} {87}},\ \bibinfo {pages} {075422}
  (\bibinfo {year} {2013})},\ \Eprint {http://arxiv.org/abs/1210.7098}
  {arXiv:1210.7098} \BibitemShut {NoStop}%
\bibitem [{\citenamefont {Irish}, \citenamefont {G\'omez-Bombarelli},\ and\
  \citenamefont {Lovett}(2014)}]{elinor1}%
  \BibitemOpen
  \bibfield  {author} {\bibinfo {author} {\bibfnamefont {E.~K.}\ \bibnamefont
  {Irish}}, \bibinfo {author} {\bibfnamefont {R.}~\bibnamefont
  {G\'omez-Bombarelli}}, \ and\ \bibinfo {author} {\bibfnamefont {B.~W.}\
  \bibnamefont {Lovett}},\ }\bibfield  {title} {\enquote {\bibinfo {title}
  {Vibration-assisted resonance in photosynthetic excitation-energy
  transfer},}\ }\href {\doibase 10.1103/PhysRevA.90.012510} {\bibfield
  {journal} {\bibinfo  {journal} {Phys. Rev. A}\ }\textbf {\bibinfo {volume}
  {90}},\ \bibinfo {pages} {012510} (\bibinfo {year} {2014})}\BibitemShut
  {NoStop}%
\bibitem [{\citenamefont {J.~O'Reilly}\ and\ \citenamefont
  {Olaya-Castro}(2014)}]{Castro2014}%
  \BibitemOpen
  \bibfield  {author} {\bibinfo {author} {\bibfnamefont {E.}~\bibnamefont
  {J.~O'Reilly}}\ and\ \bibinfo {author} {\bibfnamefont {A.}~\bibnamefont
  {Olaya-Castro}},\ }\bibfield  {title} {\enquote {\bibinfo {title}
  {Non-classicality of the molecular vibrations assisting exciton energy
  transfer at room temperature},}\ }\href {\doibase doi:10.1038/ncomms4012}
  {\bibfield  {journal} {\bibinfo  {journal} {Nature Communications}\ }\textbf
  {\bibinfo {volume} {5}} (\bibinfo {year} {2014}),\
  doi:10.1038/ncomms4012}\BibitemShut {NoStop}%
\bibitem [{\citenamefont {Killoran}, \citenamefont {Huelga},\ and\
  \citenamefont {Plenio}(2015)}]{killoran2015}%
  \BibitemOpen
  \bibfield  {author} {\bibinfo {author} {\bibfnamefont {N.}~\bibnamefont
  {Killoran}}, \bibinfo {author} {\bibfnamefont {S.~F.}\ \bibnamefont
  {Huelga}}, \ and\ \bibinfo {author} {\bibfnamefont {M.~B.}\ \bibnamefont
  {Plenio}},\ }\bibfield  {title} {\enquote {\bibinfo {title} {Enhancing
  light-harvesting power with coherent vibrational interactions: A quantum heat
  engine picture},}\ }\href {\doibase 10.1063/1.4932307} {\bibfield  {journal}
  {\bibinfo  {journal} {The Journal of Chemical Physics}\ }\textbf {\bibinfo
  {volume} {143}},\ \bibinfo {pages} {155102} (\bibinfo {year} {2015})},\
  \Eprint {http://arxiv.org/abs/http://dx.doi.org/10.1063/1.4932307}
  {http://dx.doi.org/10.1063/1.4932307} \BibitemShut {NoStop}%
\bibitem [{\citenamefont {Stones}\ and\ \citenamefont
  {Olaya-Castro}(2016)}]{Stones2016}%
  \BibitemOpen
  \bibfield  {author} {\bibinfo {author} {\bibfnamefont {R.}~\bibnamefont
  {Stones}}\ and\ \bibinfo {author} {\bibfnamefont {A.}~\bibnamefont
  {Olaya-Castro}},\ }\bibfield  {title} {\enquote {\bibinfo {title} {{Vibronic
  Coupling as a Design Principle to Optimize Photosynthetic Energy
  Transfer}},}\ }\href {\doibase 10.1016/j.chempr.2016.11.014} {\bibfield
  {journal} {\bibinfo  {journal} {Chem}\ }\textbf {\bibinfo {volume} {1}},\
  \bibinfo {pages} {822--824} (\bibinfo {year} {2016})}\BibitemShut {NoStop}%
\bibitem [{\citenamefont {Strasberg}\ \emph {et~al.}(2016)\citenamefont
  {Strasberg}, \citenamefont {Schaller}, \citenamefont {Lambert},\ and\
  \citenamefont {Brandes}}]{Strasberg_2016}%
  \BibitemOpen
  \bibfield  {author} {\bibinfo {author} {\bibfnamefont {P.}~\bibnamefont
  {Strasberg}}, \bibinfo {author} {\bibfnamefont {G.}~\bibnamefont {Schaller}},
  \bibinfo {author} {\bibfnamefont {N.}~\bibnamefont {Lambert}}, \ and\
  \bibinfo {author} {\bibfnamefont {T.}~\bibnamefont {Brandes}},\ }\bibfield
  {title} {\enquote {\bibinfo {title} {Nonequilibrium thermodynamics in the
  strong coupling and non-{Markovian} regime based on a reaction coordinate
  mapping},}\ }\href {\doibase 10.1088/1367-2630/18/7/073007} {\bibfield
  {journal} {\bibinfo  {journal} {New Journal of Physics}\ }\textbf {\bibinfo
  {volume} {18}},\ \bibinfo {pages} {073007} (\bibinfo {year}
  {2016})}\BibitemShut {NoStop}%
\bibitem [{\citenamefont {Levi}, \citenamefont {Irish},\ and\ \citenamefont
  {Lovett}(2016)}]{elinor2}%
  \BibitemOpen
  \bibfield  {author} {\bibinfo {author} {\bibfnamefont {E.~K.}\ \bibnamefont
  {Levi}}, \bibinfo {author} {\bibfnamefont {E.~K.}\ \bibnamefont {Irish}}, \
  and\ \bibinfo {author} {\bibfnamefont {B.~W.}\ \bibnamefont {Lovett}},\
  }\bibfield  {title} {\enquote {\bibinfo {title} {Coherent exciton dynamics in
  a dissipative environment maintained by an off-resonant vibrational mode},}\
  }\href {\doibase 10.1103/PhysRevA.93.042109} {\bibfield  {journal} {\bibinfo
  {journal} {Phys. Rev. A}\ }\textbf {\bibinfo {volume} {93}},\ \bibinfo
  {pages} {042109} (\bibinfo {year} {2016})}\BibitemShut {NoStop}%
\bibitem [{\citenamefont {Chen}, \citenamefont {Chiu},\ and\ \citenamefont
  {Chen}(2016)}]{Chen2016}%
  \BibitemOpen
  \bibfield  {author} {\bibinfo {author} {\bibfnamefont {H.-B.}\ \bibnamefont
  {Chen}}, \bibinfo {author} {\bibfnamefont {P.-Y.}\ \bibnamefont {Chiu}}, \
  and\ \bibinfo {author} {\bibfnamefont {Y.-N.}\ \bibnamefont {Chen}},\
  }\bibfield  {title} {\enquote {\bibinfo {title} {Vibration-induced coherence
  enhancement of the performance of a biological quantum heat engine},}\ }\href
  {\doibase 10.1103/PhysRevE.94.052101} {\bibfield  {journal} {\bibinfo
  {journal} {Phys. Rev. E}\ }\textbf {\bibinfo {volume} {94}},\ \bibinfo
  {pages} {052101} (\bibinfo {year} {2016})}\BibitemShut {NoStop}%
\bibitem [{\citenamefont {Newman}, \citenamefont {Mintert},\ and\ \citenamefont
  {Nazir}(2017)}]{Newman_2016}%
  \BibitemOpen
  \bibfield  {author} {\bibinfo {author} {\bibfnamefont {D.}~\bibnamefont
  {Newman}}, \bibinfo {author} {\bibfnamefont {F.}~\bibnamefont {Mintert}}, \
  and\ \bibinfo {author} {\bibfnamefont {A.}~\bibnamefont {Nazir}},\ }\bibfield
   {title} {\enquote {\bibinfo {title} {Performance of a quantum heat engine at
  strong reservoir coupling},}\ }\href {\doibase 10.1103/PhysRevE.95.032139}
  {\bibfield  {journal} {\bibinfo  {journal} {Phys. Rev. E}\ }\textbf {\bibinfo
  {volume} {95}},\ \bibinfo {pages} {032139} (\bibinfo {year}
  {2017})}\BibitemShut {NoStop}%
\bibitem [{\citenamefont {Qin}\ \emph {et~al.}(2017)\citenamefont {Qin},
  \citenamefont {Shen}, \citenamefont {Zhao},\ and\ \citenamefont
  {Yi}}]{Qin2017}%
  \BibitemOpen
  \bibfield  {author} {\bibinfo {author} {\bibfnamefont {M.}~\bibnamefont
  {Qin}}, \bibinfo {author} {\bibfnamefont {H.~Z.}\ \bibnamefont {Shen}},
  \bibinfo {author} {\bibfnamefont {X.~L.}\ \bibnamefont {Zhao}}, \ and\
  \bibinfo {author} {\bibfnamefont {X.~X.}\ \bibnamefont {Yi}},\ }\bibfield
  {title} {\enquote {\bibinfo {title} {Effects of system-bath coupling on a
  photosynthetic heat engine: A polaron master-equation approach},}\ }\href
  {\doibase 10.1103/PhysRevA.96.012125} {\bibfield  {journal} {\bibinfo
  {journal} {Phys. Rev. A}\ }\textbf {\bibinfo {volume} {96}},\ \bibinfo
  {pages} {012125} (\bibinfo {year} {2017})}\BibitemShut {NoStop}%
\bibitem [{\citenamefont {Br{\'e}das}, \citenamefont {Sargent},\ and\
  \citenamefont {Scholes}(2017)}]{bredas2017photovoltaic}%
  \BibitemOpen
  \bibfield  {author} {\bibinfo {author} {\bibfnamefont {J.-L.}\ \bibnamefont
  {Br{\'e}das}}, \bibinfo {author} {\bibfnamefont {E.~H.}\ \bibnamefont
  {Sargent}}, \ and\ \bibinfo {author} {\bibfnamefont {G.~D.}\ \bibnamefont
  {Scholes}},\ }\bibfield  {title} {\enquote {\bibinfo {title} {Photovoltaic
  concepts inspired by coherence effects in photosynthetic systems},}\
  }\href@noop {} {\bibfield  {journal} {\bibinfo  {journal} {Nature materials}\
  }\textbf {\bibinfo {volume} {16}},\ \bibinfo {pages} {35} (\bibinfo {year}
  {2017})}\BibitemShut {NoStop}%
\bibitem [{\citenamefont {de~Vega}\ and\ \citenamefont
  {Alonso}(2017)}]{deVega2017}%
  \BibitemOpen
  \bibfield  {author} {\bibinfo {author} {\bibfnamefont {I.}~\bibnamefont
  {de~Vega}}\ and\ \bibinfo {author} {\bibfnamefont {D.}~\bibnamefont
  {Alonso}},\ }\bibfield  {title} {\enquote {\bibinfo {title} {Dynamics of
  non-{M}arkovian open quantum systems},}\ }\href {\doibase
  10.1103/RevModPhys.89.015001} {\bibfield  {journal} {\bibinfo  {journal}
  {Rev. Mod. Phys.}\ }\textbf {\bibinfo {volume} {89}},\ \bibinfo {pages}
  {015001} (\bibinfo {year} {2017})}\BibitemShut {NoStop}%
\bibitem [{\citenamefont {Chen}\ \emph {et~al.}(2015)\citenamefont {Chen},
  \citenamefont {Lambert}, \citenamefont {Cheng}, \citenamefont {Chen},\ and\
  \citenamefont {Nori}}]{Chen2015}%
  \BibitemOpen
  \bibfield  {author} {\bibinfo {author} {\bibfnamefont {H.-B.}\ \bibnamefont
  {Chen}}, \bibinfo {author} {\bibfnamefont {N.}~\bibnamefont {Lambert}},
  \bibinfo {author} {\bibfnamefont {Y.-C.}\ \bibnamefont {Cheng}}, \bibinfo
  {author} {\bibfnamefont {Y.-N.}\ \bibnamefont {Chen}}, \ and\ \bibinfo
  {author} {\bibfnamefont {F.}~\bibnamefont {Nori}},\ }\bibfield  {title}
  {\enquote {\bibinfo {title} {{Using non-Markovian measures to evaluate
  quantum master equations for photosynthesis}},}\ }\href {\doibase
  10.1038/srep12753} {\bibfield  {journal} {\bibinfo  {journal} {Sci. Rep.}\
  }\textbf {\bibinfo {volume} {5}},\ \bibinfo {pages} {12753} (\bibinfo {year}
  {2015})},\ \Eprint {http://arxiv.org/abs/arXiv:1503.02412v1}
  {arXiv:arXiv:1503.02412v1} \BibitemShut {NoStop}%
\bibitem [{\citenamefont {Fruchtman}, \citenamefont {Lambert},\ and\
  \citenamefont {Gauger}(2016)}]{Fruchtman2016}%
  \BibitemOpen
  \bibfield  {author} {\bibinfo {author} {\bibfnamefont {A.}~\bibnamefont
  {Fruchtman}}, \bibinfo {author} {\bibfnamefont {N.}~\bibnamefont {Lambert}},
  \ and\ \bibinfo {author} {\bibfnamefont {E.~M.}\ \bibnamefont {Gauger}},\
  }\bibfield  {title} {\enquote {\bibinfo {title} {{When do perturbative
  approaches accurately capture the dynamics of complex quantum systems?}}}\
  }\href {\doibase 10.1038/srep28204} {\bibfield  {journal} {\bibinfo
  {journal} {Sci. Rep.}\ }\textbf {\bibinfo {volume} {6}},\ \bibinfo {pages}
  {28204} (\bibinfo {year} {2016})},\ \Eprint {http://arxiv.org/abs/1512.09086}
  {arXiv:1512.09086} \BibitemShut {NoStop}%
\bibitem [{\citenamefont {Tanimura}\ and\ \citenamefont
  {Kubo}(1989)}]{Tanimura_1989}%
  \BibitemOpen
  \bibfield  {author} {\bibinfo {author} {\bibfnamefont {Y.}~\bibnamefont
  {Tanimura}}\ and\ \bibinfo {author} {\bibfnamefont {R.}~\bibnamefont
  {Kubo}},\ }\bibfield  {title} {\enquote {\bibinfo {title} {Time evolution of
  a quantum system in contact with a nearly gaussian-markoffian noise bath},}\
  }\href {\doibase 10.1143/jpsj.58.101} {\bibfield  {journal} {\bibinfo
  {journal} {Journal of the Physical Society of Japan}\ }\textbf {\bibinfo
  {volume} {58}},\ \bibinfo {pages} {101--114} (\bibinfo {year}
  {1989})}\BibitemShut {NoStop}%
\bibitem [{\citenamefont {Kreisbeck}\ and\ \citenamefont
  {Kramer}(2012)}]{Kreisbeck2012}%
  \BibitemOpen
  \bibfield  {author} {\bibinfo {author} {\bibfnamefont {C.}~\bibnamefont
  {Kreisbeck}}\ and\ \bibinfo {author} {\bibfnamefont {T.}~\bibnamefont
  {Kramer}},\ }\bibfield  {title} {\enquote {\bibinfo {title} {{Long-lived
  electronic coherence in dissipative exciton dynamics of light-harvesting
  complexes}},}\ }\href {\doibase 10.1021/jz3012029} {\bibfield  {journal}
  {\bibinfo  {journal} {J. Phys. Chem. Lett.}\ }\textbf {\bibinfo {volume}
  {3}},\ \bibinfo {pages} {2828--2833} (\bibinfo {year} {2012})},\ \Eprint
  {http://arxiv.org/abs/1203.1485} {arXiv:1203.1485} \BibitemShut {NoStop}%
\bibitem [{\citenamefont {Garg}, \citenamefont {Onuchic},\ and\ \citenamefont
  {Ambegaokar}(1985)}]{Garg_1985}%
  \BibitemOpen
  \bibfield  {author} {\bibinfo {author} {\bibfnamefont {A.}~\bibnamefont
  {Garg}}, \bibinfo {author} {\bibfnamefont {J.~N.}\ \bibnamefont {Onuchic}}, \
  and\ \bibinfo {author} {\bibfnamefont {V.}~\bibnamefont {Ambegaokar}},\
  }\bibfield  {title} {\enquote {\bibinfo {title} {Effect of friction on
  electron transfer in biomolecules},}\ }\href {\doibase 10.1063/1.449017}
  {\bibfield  {journal} {\bibinfo  {journal} {The Journal of Chemical Physics}\
  }\textbf {\bibinfo {volume} {83}},\ \bibinfo {pages} {4491--4503} (\bibinfo
  {year} {1985})}\BibitemShut {NoStop}%
\bibitem [{\citenamefont {Iles-Smith}, \citenamefont {Lambert},\ and\
  \citenamefont {Nazir}(2014)}]{Iles_Smith_2014}%
  \BibitemOpen
  \bibfield  {author} {\bibinfo {author} {\bibfnamefont {J.}~\bibnamefont
  {Iles-Smith}}, \bibinfo {author} {\bibfnamefont {N.}~\bibnamefont {Lambert}},
  \ and\ \bibinfo {author} {\bibfnamefont {A.}~\bibnamefont {Nazir}},\
  }\bibfield  {title} {\enquote {\bibinfo {title} {Environmental dynamics,
  correlations, and the emergence of noncanonical equilibrium states in open
  quantum systems},}\ }\href {\doibase 10.1103/physreva.90.032114} {\bibfield
  {journal} {\bibinfo  {journal} {Physical Review A}\ }\textbf {\bibinfo
  {volume} {90}} (\bibinfo {year} {2014}),\
  10.1103/physreva.90.032114}\BibitemShut {NoStop}%
\bibitem [{\citenamefont {Iles-Smith}\ \emph {et~al.}(2016)\citenamefont
  {Iles-Smith}, \citenamefont {Dijkstra}, \citenamefont {Lambert},\ and\
  \citenamefont {Nazir}}]{Iles_Smith_2015}%
  \BibitemOpen
  \bibfield  {author} {\bibinfo {author} {\bibfnamefont {J.}~\bibnamefont
  {Iles-Smith}}, \bibinfo {author} {\bibfnamefont {A.~G.}\ \bibnamefont
  {Dijkstra}}, \bibinfo {author} {\bibfnamefont {N.}~\bibnamefont {Lambert}}, \
  and\ \bibinfo {author} {\bibfnamefont {A.}~\bibnamefont {Nazir}},\ }\bibfield
   {title} {\enquote {\bibinfo {title} {Energy transfer in structured and
  unstructured environments: master equations beyond the born-markov
  approximations},}\ }\href {\doibase 10.1063/1.4940218} {\bibfield  {journal}
  {\bibinfo  {journal} {J. Chem. Phys.}\ }\textbf {\bibinfo {volume} {144}},\
  \bibinfo {pages} {044110} (\bibinfo {year} {2016})}\BibitemShut {NoStop}%
\bibitem [{\citenamefont {Makri}(1998)}]{Makri_1998}%
  \BibitemOpen
  \bibfield  {author} {\bibinfo {author} {\bibfnamefont {N.}~\bibnamefont
  {Makri}},\ }\bibfield  {title} {\enquote {\bibinfo {title} {Quantum
  dissipative dynamics:~ a numerically exact methodology},}\ }\href {\doibase
  10.1021/jp980359y} {\bibfield  {journal} {\bibinfo  {journal} {The Journal of
  Physical Chemistry A}\ }\textbf {\bibinfo {volume} {102}},\ \bibinfo {pages}
  {4414--4427} (\bibinfo {year} {1998})}\BibitemShut {NoStop}%
\bibitem [{\citenamefont {Nalbach}\ \emph {et~al.}(2011)\citenamefont
  {Nalbach}, \citenamefont {Ishizaki}, \citenamefont {Fleming},\ and\
  \citenamefont {Thorwart}}]{Peter11}%
  \BibitemOpen
  \bibfield  {author} {\bibinfo {author} {\bibfnamefont {P.}~\bibnamefont
  {Nalbach}}, \bibinfo {author} {\bibfnamefont {A.}~\bibnamefont {Ishizaki}},
  \bibinfo {author} {\bibfnamefont {G.~R.}\ \bibnamefont {Fleming}}, \ and\
  \bibinfo {author} {\bibfnamefont {M.}~\bibnamefont {Thorwart}},\ }\bibfield
  {title} {\enquote {\bibinfo {title} {{Iterative path-integral algorithm
  versus cumulant time-nonlocal master equation approach for dissipative
  biomolecular exciton transport}},}\ }\href {\doibase
  10.1088/1367-2630/13/6/063040} {\bibfield  {journal} {\bibinfo  {journal}
  {New J. Phys.}\ }\textbf {\bibinfo {volume} {13}},\ \bibinfo {pages} {063040}
  (\bibinfo {year} {2011})}\BibitemShut {NoStop}%
\bibitem [{\citenamefont {Manthe}(2008)}]{manthe2008multilayer}%
  \BibitemOpen
  \bibfield  {author} {\bibinfo {author} {\bibfnamefont {U.}~\bibnamefont
  {Manthe}},\ }\bibfield  {title} {\enquote {\bibinfo {title} {A multilayer
  multiconfigurational time-dependent hartree approach for quantum dynamics on
  general potential energy surfaces},}\ }\href@noop {} {\bibfield  {journal}
  {\bibinfo  {journal} {The Journal of chemical physics}\ }\textbf {\bibinfo
  {volume} {128}},\ \bibinfo {pages} {164116} (\bibinfo {year}
  {2008})}\BibitemShut {NoStop}%
\bibitem [{\citenamefont {Schulze}\ and\ \citenamefont
  {Kuhn}(2015)}]{schulze2015explicit}%
  \BibitemOpen
  \bibfield  {author} {\bibinfo {author} {\bibfnamefont {J.}~\bibnamefont
  {Schulze}}\ and\ \bibinfo {author} {\bibfnamefont {O.}~\bibnamefont {Kuhn}},\
  }\bibfield  {title} {\enquote {\bibinfo {title} {Explicit correlated
  exciton-vibrational dynamics of the fmo complex},}\ }\href@noop {} {\bibfield
   {journal} {\bibinfo  {journal} {The Journal of Physical Chemistry B}\
  }\textbf {\bibinfo {volume} {119}},\ \bibinfo {pages} {6211--6216} (\bibinfo
  {year} {2015})}\BibitemShut {NoStop}%
\bibitem [{\citenamefont {Chin}\ \emph
  {et~al.}(2013{\natexlab{b}})\citenamefont {Chin}, \citenamefont {Prior},
  \citenamefont {Rosenbach}, \citenamefont {Caycedo-Soler}, \citenamefont
  {Huelga},\ and\ \citenamefont {Plenio}}]{chin2013role}%
  \BibitemOpen
  \bibfield  {author} {\bibinfo {author} {\bibfnamefont {A.}~\bibnamefont
  {Chin}}, \bibinfo {author} {\bibfnamefont {J.}~\bibnamefont {Prior}},
  \bibinfo {author} {\bibfnamefont {R.}~\bibnamefont {Rosenbach}}, \bibinfo
  {author} {\bibfnamefont {F.}~\bibnamefont {Caycedo-Soler}}, \bibinfo {author}
  {\bibfnamefont {S.}~\bibnamefont {Huelga}}, \ and\ \bibinfo {author}
  {\bibfnamefont {M.}~\bibnamefont {Plenio}},\ }\bibfield  {title} {\enquote
  {\bibinfo {title} {The role of non-equilibrium vibrational structures in
  electronic coherence and recoherence in pigment--protein complexes},}\
  }\href@noop {} {\bibfield  {journal} {\bibinfo  {journal} {Nature Physics}\
  }\textbf {\bibinfo {volume} {9}},\ \bibinfo {pages} {113} (\bibinfo {year}
  {2013}{\natexlab{b}})}\BibitemShut {NoStop}%
\bibitem [{\citenamefont {Prior}\ \emph {et~al.}(2013)\citenamefont {Prior},
  \citenamefont {de~Vega}, \citenamefont {Chin}, \citenamefont {Huelga},\ and\
  \citenamefont {Plenio}}]{prior2013quantum}%
  \BibitemOpen
  \bibfield  {author} {\bibinfo {author} {\bibfnamefont {J.}~\bibnamefont
  {Prior}}, \bibinfo {author} {\bibfnamefont {I.}~\bibnamefont {de~Vega}},
  \bibinfo {author} {\bibfnamefont {A.~W.}\ \bibnamefont {Chin}}, \bibinfo
  {author} {\bibfnamefont {S.~F.}\ \bibnamefont {Huelga}}, \ and\ \bibinfo
  {author} {\bibfnamefont {M.~B.}\ \bibnamefont {Plenio}},\ }\bibfield  {title}
  {\enquote {\bibinfo {title} {Quantum dynamics in photonic crystals},}\
  }\href@noop {} {\bibfield  {journal} {\bibinfo  {journal} {Physical Review
  A}\ }\textbf {\bibinfo {volume} {87}},\ \bibinfo {pages} {013428} (\bibinfo
  {year} {2013})}\BibitemShut {NoStop}%
\bibitem [{\citenamefont {Schr{\"o}der}\ \emph {et~al.}(2017)\citenamefont
  {Schr{\"o}der}, \citenamefont {Turban}, \citenamefont {Musser}, \citenamefont
  {Hine},\ and\ \citenamefont {Chin}}]{schroder2017multi}%
  \BibitemOpen
  \bibfield  {author} {\bibinfo {author} {\bibfnamefont {F.~A. Y.~N.}\
  \bibnamefont {Schr{\"o}der}}, \bibinfo {author} {\bibfnamefont {D.~H.~P.}\
  \bibnamefont {Turban}}, \bibinfo {author} {\bibfnamefont {A.~J.}\
  \bibnamefont {Musser}}, \bibinfo {author} {\bibfnamefont {N.~D.~M.}\
  \bibnamefont {Hine}}, \ and\ \bibinfo {author} {\bibfnamefont {A.~W.}\
  \bibnamefont {Chin}},\ }\bibfield  {title} {\enquote {\bibinfo {title}
  {Multi-dimensional tensor network simulation of open quantum dynamics in
  singlet fission},}\ }\href@noop {} {\bibfield  {journal} {\bibinfo  {journal}
  {arXiv preprint arXiv:1710.01362}\ } (\bibinfo {year} {2017})}\BibitemShut
  {NoStop}%
\bibitem [{\citenamefont {Fujihashi}\ \emph {et~al.}(2017)\citenamefont
  {Fujihashi}, \citenamefont {Chen}, \citenamefont {Ishizaki}, \citenamefont
  {Wang},\ and\ \citenamefont {Zhao}}]{Ishizaki_2017}%
  \BibitemOpen
  \bibfield  {author} {\bibinfo {author} {\bibfnamefont {Y.}~\bibnamefont
  {Fujihashi}}, \bibinfo {author} {\bibfnamefont {L.}~\bibnamefont {Chen}},
  \bibinfo {author} {\bibfnamefont {A.}~\bibnamefont {Ishizaki}}, \bibinfo
  {author} {\bibfnamefont {J.}~\bibnamefont {Wang}}, \ and\ \bibinfo {author}
  {\bibfnamefont {Y.}~\bibnamefont {Zhao}},\ }\bibfield  {title} {\enquote
  {\bibinfo {title} {Effect of high-frequency modes on singlet fission
  dynamics},}\ }\href {\doibase 10.1063/1.4973981} {\bibfield  {journal}
  {\bibinfo  {journal} {The Journal of Chemical Physics}\ }\textbf {\bibinfo
  {volume} {146}},\ \bibinfo {pages} {044101} (\bibinfo {year} {2017})},\
  \Eprint {http://arxiv.org/abs/http://dx.doi.org/10.1063/1.4973981}
  {http://dx.doi.org/10.1063/1.4973981} \BibitemShut {NoStop}%
\bibitem [{\citenamefont {Stones}\ \emph {et~al.}()\citenamefont {Stones},
  \citenamefont {Hossein-Nejad}, \citenamefont {van Grondelle},\ and\
  \citenamefont {Olaya-Castro}}]{Stones2016a}%
  \BibitemOpen
  \bibfield  {author} {\bibinfo {author} {\bibfnamefont {R.}~\bibnamefont
  {Stones}}, \bibinfo {author} {\bibfnamefont {H.}~\bibnamefont
  {Hossein-Nejad}}, \bibinfo {author} {\bibfnamefont {R.}~\bibnamefont {van
  Grondelle}}, \ and\ \bibinfo {author} {\bibfnamefont {A.}~\bibnamefont
  {Olaya-Castro}},\ }\bibfield  {title} {\enquote {\bibinfo {title} {On the
  perfomance of a photosystem {II} reaction centre-based photocell},}\
  }\href@noop {} {\ }\Eprint {http://arxiv.org/abs/1601.05260v2} {1601.05260v2}
  \BibitemShut {NoStop}%
\bibitem [{\citenamefont {Giusteri}\ \emph {et~al.}(2017)\citenamefont
  {Giusteri}, \citenamefont {Recrosi}, \citenamefont {Schaller},\ and\
  \citenamefont {Celardo}}]{Giusteri2017}%
  \BibitemOpen
  \bibfield  {author} {\bibinfo {author} {\bibfnamefont {G.~G.}\ \bibnamefont
  {Giusteri}}, \bibinfo {author} {\bibfnamefont {F.}~\bibnamefont {Recrosi}},
  \bibinfo {author} {\bibfnamefont {G.}~\bibnamefont {Schaller}}, \ and\
  \bibinfo {author} {\bibfnamefont {G.~L.}\ \bibnamefont {Celardo}},\
  }\bibfield  {title} {\enquote {\bibinfo {title} {Interplay of different
  environments in open quantum systems: Breakdown of the additive
  approximation},}\ }\href {\doibase 10.1103/physreve.96.012113} {\bibfield
  {journal} {\bibinfo  {journal} {Physical Review E}\ }\textbf {\bibinfo
  {volume} {96}} (\bibinfo {year} {2017}),\
  10.1103/physreve.96.012113}\BibitemShut {NoStop}%
\bibitem [{\citenamefont {Gruss}, \citenamefont {Smolyanitsky},\ and\
  \citenamefont {Zwolak}(2017)}]{gruss2017communication}%
  \BibitemOpen
  \bibfield  {author} {\bibinfo {author} {\bibfnamefont {D.}~\bibnamefont
  {Gruss}}, \bibinfo {author} {\bibfnamefont {A.}~\bibnamefont {Smolyanitsky}},
  \ and\ \bibinfo {author} {\bibfnamefont {M.}~\bibnamefont {Zwolak}},\
  }\bibfield  {title} {\enquote {\bibinfo {title} {Communication:
  Relaxation-limited electronic currents in extended reservoir simulations},}\
  }\href@noop {} {\bibfield  {journal} {\bibinfo  {journal} {The Journal of
  Chemical Physics}\ }\textbf {\bibinfo {volume} {147}},\ \bibinfo {pages}
  {141102} (\bibinfo {year} {2017})}\BibitemShut {NoStop}%
\bibitem [{\citenamefont {Ghosh}, \citenamefont {Smirnov},\ and\ \citenamefont
  {Nori}(2009)}]{Ghosh2009}%
  \BibitemOpen
  \bibfield  {author} {\bibinfo {author} {\bibfnamefont {P.~K.}\ \bibnamefont
  {Ghosh}}, \bibinfo {author} {\bibfnamefont {A.~Y.}\ \bibnamefont {Smirnov}},
  \ and\ \bibinfo {author} {\bibfnamefont {F.}~\bibnamefont {Nori}},\
  }\bibfield  {title} {\enquote {\bibinfo {title} {{Modeling light-driven
  proton pumps in artificial photosynthetic reaction centers}},}\ }\href
  {\doibase 10.1063/1.3170939} {\bibfield  {journal} {\bibinfo  {journal} {J.
  Chem. Phys.}\ }\textbf {\bibinfo {volume} {131}},\ \bibinfo {pages} {035102}
  (\bibinfo {year} {2009})},\ \Eprint {http://arxiv.org/abs/arXiv:0901.2170v2}
  {arXiv:arXiv:0901.2170v2} \BibitemShut {NoStop}%
\bibitem [{\citenamefont {Dodin}, \citenamefont {Tscherbul},\ and\
  \citenamefont {Brumer}(2016)}]{brumer2016}%
  \BibitemOpen
  \bibfield  {author} {\bibinfo {author} {\bibfnamefont {A.}~\bibnamefont
  {Dodin}}, \bibinfo {author} {\bibfnamefont {T.~V.}\ \bibnamefont
  {Tscherbul}}, \ and\ \bibinfo {author} {\bibfnamefont {P.}~\bibnamefont
  {Brumer}},\ }\bibfield  {title} {\enquote {\bibinfo {title} {Quantum dynamics
  of incoherently driven v-type systems: Analytic solutions beyond the secular
  approximation},}\ }\href {\doibase 10.1063/1.4954243} {\bibfield  {journal}
  {\bibinfo  {journal} {The Journal of Chemical Physics}\ }\textbf {\bibinfo
  {volume} {144}},\ \bibinfo {pages} {244108} (\bibinfo {year} {2016})},\
  \Eprint {http://arxiv.org/abs/http://dx.doi.org/10.1063/1.4954243}
  {http://dx.doi.org/10.1063/1.4954243} \BibitemShut {NoStop}%
\bibitem [{\citenamefont {Wendling}\ \emph {et~al.}(2000)\citenamefont
  {Wendling}, \citenamefont {Pullerits}, \citenamefont {Przyjalgowski},
  \citenamefont {Vulto}, \citenamefont {Aartsma}, \citenamefont {van
  Grondelle},\ and\ \citenamefont {van Amerongen}}]{wendling}%
  \BibitemOpen
  \bibfield  {author} {\bibinfo {author} {\bibfnamefont {M.}~\bibnamefont
  {Wendling}}, \bibinfo {author} {\bibfnamefont {T.}~\bibnamefont {Pullerits}},
  \bibinfo {author} {\bibfnamefont {M.~A.}\ \bibnamefont {Przyjalgowski}},
  \bibinfo {author} {\bibfnamefont {S.~I.~E.}\ \bibnamefont {Vulto}}, \bibinfo
  {author} {\bibfnamefont {T.~J.}\ \bibnamefont {Aartsma}}, \bibinfo {author}
  {\bibfnamefont {R.}~\bibnamefont {van Grondelle}}, \ and\ \bibinfo {author}
  {\bibfnamefont {H.}~\bibnamefont {van Amerongen}},\ }\bibfield  {title}
  {\enquote {\bibinfo {title} {Electron-vibrational coupling in the
  fenna-matthews-olson complex of prosthecochloris aestuarii determined by
  temperature-dependent absorption and fluorescence line-narrowing
  measurements},}\ }\href {\doibase 10.1021/jp000077+} {\bibfield  {journal}
  {\bibinfo  {journal} {The Journal of Physical Chemistry B}\ }\textbf
  {\bibinfo {volume} {104}},\ \bibinfo {pages} {5825--5831} (\bibinfo {year}
  {2000})},\ \Eprint {http://arxiv.org/abs/http://dx.doi.org/10.1021/jp000077+}
  {http://dx.doi.org/10.1021/jp000077+} \BibitemShut {NoStop}%
\bibitem [{\citenamefont {Martinazzo}\ \emph {et~al.}(2011)\citenamefont
  {Martinazzo}, \citenamefont {Vacchini}, \citenamefont {Hughes},\ and\
  \citenamefont {Burghardt}}]{Martinazzo_2011}%
  \BibitemOpen
  \bibfield  {author} {\bibinfo {author} {\bibfnamefont {R.}~\bibnamefont
  {Martinazzo}}, \bibinfo {author} {\bibfnamefont {B.}~\bibnamefont
  {Vacchini}}, \bibinfo {author} {\bibfnamefont {K.~H.}\ \bibnamefont
  {Hughes}}, \ and\ \bibinfo {author} {\bibfnamefont {I.}~\bibnamefont
  {Burghardt}},\ }\bibfield  {title} {\enquote {\bibinfo {title}
  {Communication: Universal {M}arkovian reduction of {B}rownian particle
  dynamics},}\ }\href {\doibase 10.1063/1.3532408} {\bibfield  {journal}
  {\bibinfo  {journal} {The Journal of Chemical Physics}\ }\textbf {\bibinfo
  {volume} {134}},\ \bibinfo {pages} {011101} (\bibinfo {year}
  {2011})}\BibitemShut {NoStop}%
\bibitem [{\citenamefont {Johansson}, \citenamefont {Nation},\ and\
  \citenamefont {Nori}(2012)}]{qutip}%
  \BibitemOpen
  \bibfield  {author} {\bibinfo {author} {\bibfnamefont {J.~R.}\ \bibnamefont
  {Johansson}}, \bibinfo {author} {\bibfnamefont {P.~D.}\ \bibnamefont
  {Nation}}, \ and\ \bibinfo {author} {\bibfnamefont {F.}~\bibnamefont
  {Nori}},\ }\bibfield  {title} {\enquote {\bibinfo {title} {Qutip: An
  open-source python framework for the dynamics of open quantum systems},}\
  }\href {\doibase 10.1016/j.cpc.2012.02.021} {\bibfield  {journal} {\bibinfo
  {journal} {Computer Physics Communications}\ } (\bibinfo {year} {2012}),\
  10.1016/j.cpc.2012.02.021}\BibitemShut {NoStop}%
\bibitem [{\citenamefont {Johansson}, \citenamefont {Nation},\ and\
  \citenamefont {Nori}(2013)}]{qutip2}%
  \BibitemOpen
  \bibfield  {author} {\bibinfo {author} {\bibfnamefont {J.~R.}\ \bibnamefont
  {Johansson}}, \bibinfo {author} {\bibfnamefont {P.~D.}\ \bibnamefont
  {Nation}}, \ and\ \bibinfo {author} {\bibfnamefont {F.}~\bibnamefont
  {Nori}},\ }\bibfield  {title} {\enquote {\bibinfo {title} {{QuTiP} 2: A
  python framework for the dynamics of open quantum systems},}\ }\href
  {\doibase 10.1016/j.cpc.2012.11.019} {\bibfield  {journal} {\bibinfo
  {journal} {Computer Physics Communications}\ }\textbf {\bibinfo {volume}
  {184}},\ \bibinfo {pages} {1234--1240} (\bibinfo {year} {2013})}\BibitemShut
  {NoStop}%
\bibitem [{\citenamefont {Breuer}\ and\ \citenamefont
  {Petruccione}(2002)}]{Breuer2002}%
  \BibitemOpen
  \bibfield  {author} {\bibinfo {author} {\bibfnamefont {H.-P.}\ \bibnamefont
  {Breuer}}\ and\ \bibinfo {author} {\bibfnamefont {F.}~\bibnamefont
  {Petruccione}},\ }\href@noop {} {\emph {\bibinfo {title} {The Theory of Open
  Quantum Systems}}}\ (\bibinfo  {publisher} {Oxford University Press},\
  \bibinfo {year} {2002})\BibitemShut {NoStop}%
\bibitem [{\citenamefont {Ashhab}\ \emph {et~al.}(2009)\citenamefont {Ashhab},
  \citenamefont {Johansson}, \citenamefont {Zagoskin},\ and\ \citenamefont
  {Nori}}]{ashhab}%
  \BibitemOpen
  \bibfield  {author} {\bibinfo {author} {\bibfnamefont {S.}~\bibnamefont
  {Ashhab}}, \bibinfo {author} {\bibfnamefont {J.~R.}\ \bibnamefont
  {Johansson}}, \bibinfo {author} {\bibfnamefont {A.~M.}\ \bibnamefont
  {Zagoskin}}, \ and\ \bibinfo {author} {\bibfnamefont {F.}~\bibnamefont
  {Nori}},\ }\bibfield  {title} {\enquote {\bibinfo {title}
  {Single-artificial-atom lasing using a voltage-biased superconducting charge
  qubit},}\ }\href {http://stacks.iop.org/1367-2630/11/i=2/a=023030} {\bibfield
   {journal} {\bibinfo  {journal} {New Journal of Physics}\ }\textbf {\bibinfo
  {volume} {11}},\ \bibinfo {pages} {023030} (\bibinfo {year}
  {2009})}\BibitemShut {NoStop}%
\bibitem [{\citenamefont {Fruchtman}\ \emph {et~al.}(2016)\citenamefont
  {Fruchtman}, \citenamefont {G\'omez-Bombarelli}, \citenamefont {Lovett},\
  and\ \citenamefont {Gauger}}]{amir2015}%
  \BibitemOpen
  \bibfield  {author} {\bibinfo {author} {\bibfnamefont {A.}~\bibnamefont
  {Fruchtman}}, \bibinfo {author} {\bibfnamefont {R.}~\bibnamefont
  {G\'omez-Bombarelli}}, \bibinfo {author} {\bibfnamefont {B.~W.}\ \bibnamefont
  {Lovett}}, \ and\ \bibinfo {author} {\bibfnamefont {E.~M.}\ \bibnamefont
  {Gauger}},\ }\bibfield  {title} {\enquote {\bibinfo {title} {Photocell
  optimization using dark state protection},}\ }\href {\doibase
  10.1103/PhysRevLett.117.203603} {\bibfield  {journal} {\bibinfo  {journal}
  {Phys. Rev. Lett.}\ }\textbf {\bibinfo {volume} {117}},\ \bibinfo {pages}
  {203603} (\bibinfo {year} {2016})}\BibitemShut {NoStop}%
\bibitem [{\citenamefont {Potocnik}\ \emph {et~al.}(2018)\citenamefont
  {Potocnik}, \citenamefont {Bargerbos}, \citenamefont {Schr{\"o}der},
  \citenamefont {Khan}, \citenamefont {Collodo}, \citenamefont {Gasparinetti},
  \citenamefont {Salathe}, \citenamefont {Creatore}, \citenamefont {Eichler},
  \citenamefont {T{\"o}reci}, \citenamefont {Chin},\ and\ \citenamefont
  {Wallraff}}]{anton}%
  \BibitemOpen
  \bibfield  {author} {\bibinfo {author} {\bibfnamefont {A.}~\bibnamefont
  {Potocnik}}, \bibinfo {author} {\bibfnamefont {A.}~\bibnamefont {Bargerbos}},
  \bibinfo {author} {\bibfnamefont {F.~A. Y.~N.}\ \bibnamefont {Schr{\"o}der}},
  \bibinfo {author} {\bibfnamefont {S.~A.}\ \bibnamefont {Khan}}, \bibinfo
  {author} {\bibfnamefont {M.~C.}\ \bibnamefont {Collodo}}, \bibinfo {author}
  {\bibfnamefont {S.}~\bibnamefont {Gasparinetti}}, \bibinfo {author}
  {\bibfnamefont {Y.}~\bibnamefont {Salathe}}, \bibinfo {author} {\bibfnamefont
  {C.}~\bibnamefont {Creatore}}, \bibinfo {author} {\bibfnamefont
  {C.}~\bibnamefont {Eichler}}, \bibinfo {author} {\bibfnamefont {H.~E.}\
  \bibnamefont {T{\"o}reci}}, \bibinfo {author} {\bibfnamefont {A.~W.}\
  \bibnamefont {Chin}}, \ and\ \bibinfo {author} {\bibfnamefont
  {A.}~\bibnamefont {Wallraff}},\ }\bibfield  {title} {\enquote {\bibinfo
  {title} {Studying light-harvesting models with superconducting circuits},}\
  }\href@noop {} {\bibfield  {journal} {\bibinfo  {journal} {Nature
  Communications}\ }\textbf {\bibinfo {volume} {9}},\ \bibinfo {pages} {904}
  (\bibinfo {year} {2018})}\BibitemShut {NoStop}%
\bibitem [{\citenamefont {Gorman}\ \emph {et~al.}(2018)\citenamefont {Gorman},
  \citenamefont {Hemmerling}, \citenamefont {Megidish}, \citenamefont
  {Moeller}, \citenamefont {Schindler}, \citenamefont {Sarovar},\ and\
  \citenamefont {Haeffner}}]{PRX}%
  \BibitemOpen
  \bibfield  {author} {\bibinfo {author} {\bibfnamefont {D.~J.}\ \bibnamefont
  {Gorman}}, \bibinfo {author} {\bibfnamefont {B.}~\bibnamefont {Hemmerling}},
  \bibinfo {author} {\bibfnamefont {E.}~\bibnamefont {Megidish}}, \bibinfo
  {author} {\bibfnamefont {S.~A.}\ \bibnamefont {Moeller}}, \bibinfo {author}
  {\bibfnamefont {P.}~\bibnamefont {Schindler}}, \bibinfo {author}
  {\bibfnamefont {M.}~\bibnamefont {Sarovar}}, \ and\ \bibinfo {author}
  {\bibfnamefont {H.}~\bibnamefont {Haeffner}},\ }\bibfield  {title} {\enquote
  {\bibinfo {title} {Engineering vibrationally assisted energy transfer in a
  trapped-ion quantum simulator},}\ }\href {\doibase 10.1103/PhysRevX.8.011038}
  {\bibfield  {journal} {\bibinfo  {journal} {Phys. Rev. X}\ }\textbf {\bibinfo
  {volume} {8}},\ \bibinfo {pages} {011038} (\bibinfo {year}
  {2018})}\BibitemShut {NoStop}%
\bibitem [{\citenamefont {Wang}\ \emph {et~al.}(2018)\citenamefont {Wang},
  \citenamefont {Tao}, \citenamefont {Ai}, \citenamefont {Xin}, \citenamefont
  {Lambert}, \citenamefont {Ruan}, \citenamefont {Cheng}, \citenamefont {Nori},
  \citenamefont {Deng},\ and\ \citenamefont {Long}}]{neill}%
  \BibitemOpen
  \bibfield  {author} {\bibinfo {author} {\bibfnamefont {B.-X.}\ \bibnamefont
  {Wang}}, \bibinfo {author} {\bibfnamefont {M.-J.}\ \bibnamefont {Tao}},
  \bibinfo {author} {\bibfnamefont {Q.}~\bibnamefont {Ai}}, \bibinfo {author}
  {\bibfnamefont {T.}~\bibnamefont {Xin}}, \bibinfo {author} {\bibfnamefont
  {N.}~\bibnamefont {Lambert}}, \bibinfo {author} {\bibfnamefont
  {D.}~\bibnamefont {Ruan}}, \bibinfo {author} {\bibfnamefont {Y.-C.}\
  \bibnamefont {Cheng}}, \bibinfo {author} {\bibfnamefont {F.}~\bibnamefont
  {Nori}}, \bibinfo {author} {\bibfnamefont {F.-G.}\ \bibnamefont {Deng}}, \
  and\ \bibinfo {author} {\bibfnamefont {G.-L.}\ \bibnamefont {Long}},\
  }\bibfield  {title} {\enquote {\bibinfo {title} {Quantum simulation of
  photosynthetic energy transfer},}\ }\href@noop {} {\bibfield  {journal}
  {\bibinfo  {journal} {arXiv:1801.09475}\ } (\bibinfo {year}
  {2018})}\BibitemShut {NoStop}%
\end{thebibliography}%
\end{document}